\PassOptionsToPackage{prologue,dvipsnames}{xcolor}
\documentclass[sigconf,authorversion]{acmart} 

\AtBeginDocument{%
  }

\copyrightyear{2025}
\acmYear{2025}
\setcopyright{none} 
\setcctype{by}
\acmConference[CHI '25]{CHI Conference on Human Factors in Computing Systems}{April 26-May 1, 2025}{Yokohama, Japan}
\acmBooktitle{CHI Conference on Human Factors in Computing Systems (CHI '25), April 26-May 1, 2025, Yokohama, Japan}\acmDOI{10.1145/3706598.3714020}
\acmISBN{979-8-4007-1394-1/25/04}

\usepackage{enumitem}
\usepackage{multirow}
\usepackage{subcaption}
\usepackage{graphicx}
\usepackage{amssymb}
\usepackage{pifont}
\newcommand{\cmark}{\ding{51}}%
\newcommand{\xmark}{\ding{55}}%

\newcommand{\var}[1]{\texttt{#1}}

\usepackage[capitalize]{cleveref}
\crefname{section}{Section}{Sections}
\Crefname{section}{Section}{Sections}
\Crefname{table}{Table}{Tables}
\crefname{table}{Table}{Tables}
\crefname{figure}{Figure}{Figures}

\newcommand{\interviewprefix}{P}
\makeatletter
\newcommand{\interview}[1]{%
  \def\nextitem{\def\nextitem{, }}
  (\@for\el:=#1\do{\nextitem\interviewprefix\el})%
}
\newcommand{\shortquote}[1]{“\textit{#1}”}%
\makeatother%

\begin{document}

\title[Fostering Appropriate Reliance on Large Language Models]{Fostering Appropriate Reliance on Large Language Models:\\The Role of Explanations, Sources, and Inconsistencies}

\author{Sunnie S. Y. Kim}
\email{sunniesuhyoung@princeton.edu}
\affiliation{%
  \institution{Princeton University}
  \state{New Jersey}
  \country{USA}
}

\author{Jennifer Wortman Vaughan}
\email{jenn@microsoft.com}
\affiliation{%
  \institution{Microsoft Research}
  \state{New York}
  \country{USA}
}

\author{Q. Vera Liao}
\email{veraliao@microsoft.com}
\affiliation{%
  \institution{Microsoft Research}
  \state{Montreal}
  \country{Canada}
}

\author{Tania Lombrozo}
\email{lombrozo@princeton.edu}
\affiliation{%
  \institution{Princeton University}
  \state{New Jersey}
  \country{USA}
}

\author{Olga Russakovsky}
\email{olgarus@princeton.edu}
\affiliation{%
  \institution{Princeton University}
  \state{New Jersey}
  \country{USA}
}

\renewcommand{\shortauthors}{Kim, Vaughan, Liao, Lombrozo, Russakovsky}

\begin{abstract} 
Large language models (LLMs) can produce erroneous responses that sound fluent and convincing, raising the risk that users will rely on these responses as if they were correct. Mitigating such overreliance is a key challenge. 
Through a think-aloud study in which participants use an LLM-infused application to answer objective questions, we identify several features of LLM responses that shape users' reliance: \textit{explanations} (supporting details for answers), \textit{inconsistencies} in explanations, and \textit{sources}.
Through a large-scale, pre-registered, controlled experiment (N=308), we isolate and study the effects of these features on users' reliance, accuracy, and other measures.
We find that the presence of explanations increases reliance on both correct and incorrect responses. However, we observe less reliance on incorrect responses when sources are provided or when explanations exhibit inconsistencies. We discuss the implications of these findings for fostering appropriate reliance on LLMs.
\end{abstract}



\maketitle

\section{Introduction}
\label{sec:intro}

Large language models (LLMs) are powerful tools, capable of a wide range of tasks from text summarization to sentence completion to code generation. Technology companies have leapt at the unprecedented opportunity to build LLM-infused applications that help users with information retrieval and search, learning new things, and performing everyday tasks more efficiently.
Many such applications, such as LLM-infused search engines and chatbots, are predicated on LLMs' ability to provide intricate responses to complex user questions. 
Already millions of people use LLMs to find answers to their questions about health, science, current events, and other domains, and the use of LLMs is widely predicted to grow~\cite{Sharma2024Echochamber,Zhu2023LargeLM,Kapoor2024ICML}. 
However, the responses produced by LLMs are often inaccurate, sometimes in subtle ways~\cite{Ji2023Hallucination,shuster2021eval,santhanam2022rome,Dahl2024Law}.
Inaccurate LLM responses have the potential to mislead users, raising the risk that users will take actions based on the assumption that responses are correct~\cite{Kim2024FAccT,Vasconcelos2023CSCW,nytimes2023lawyer,passi2024appropriate,Weidinger2022Risk}.
While such \emph{overreliance} on AI systems is not a new problem~\cite{zhang2020effect,Bansal2021CHI,Poursabzi-Sangdeh-CHI2021,wang2021explanations,passi2022overreliance}, it may be exacerbated by the introduction of LLMs, since LLM responses are often fluent and convincing even when wrong and public excitement around LLMs is high.

When asked to answer a question, LLMs and systems based on them typically provide a response that contains both an answer to the question and some supporting details or justification for this answer~\cite{Lee2024FAccT,lfqa23}. For example, when asked a math question, an LLM may provide a step-by-step derivation for its answer \cite{Collins2024PNAS,hendrycks2021measuring}. In line with everyday usage and much of the psychology literature~\cite{Lombrozo2012,Lombrozo2006TCS,keil2006explanation}, we refer to such supporting details as an \emph{explanation} of the answer.
(We note that this differs from how the term explanation is often used within the explainable AI community in that we do not make any assumptions about the extent to which it faithfully describes the way that the model arrived at its answer. That is, the explanation describes why the answer is correct, not necessarily why the model output the answer that it did.)
Some authors have argued that such explanations should help users spot incorrect answers, potentially mitigating overreliance~\cite{Gonzalez2021DoEH,Bussone2015,Lai2019FAccT,Vasconcelos2023CSCW}. However, prior work suggests that in many settings, the very presence of an explanation can increase trust and reliance, whether or not it is warranted~\cite{Bansal2021CHI,zhang2020effect,Poursabzi-Sangdeh-CHI2021,wang2021explanations,Fok2024Verifiability,Pafla2024CHI}.
To avoid such unintended negative consequences, it is necessary to understand how users interpret and act upon explanations from LLMs, and how explanations and other features of LLM responses might be adjusted to encourage appropriate reliance.

To explore these questions, we first conduct a think-aloud study with 16 participants with varying knowledge of and experience with LLMs. In this study, participants answer objective questions with the use of the popular LLM-infused application ChatGPT via multi-turn interactions. The goal of this preliminary study is to understand how people perceive LLM responses and which features of a response shape their reliance.
We observe that participants interpret \textit{inconsistencies} in explanations --- that is, sets of statements that cannot be true at the same time~\cite{Hurley2000} --- as a cue of unreliability.
Participants also seek out \textit{sources} to verify supporting details in LLM responses and are less likely to rely on incorrect answers when the sources provide helpful information.

Building on the findings from this study, we next conduct a large-scale, pre-registered, controlled experiment ($N=308$) in which participants answer difficult objective questions with access to LLM responses, i.e., responses from a hypothetical LLM named ``Theta.''\footnote{We note that the line between what we would call an ``LLM'' as opposed to an ``LLM-infused system'' can be blurry, especially when the system takes the form of a chatbot such as Theta or ChatGPT. Throughout the paper, use the term LLM for readability in places where the distinction is not important.}
These responses were created in advance using state-of-the-art LLM-infused applications ChatGPT and Perplexity AI so that we can fully control their features. Specifically, we employ a 2 x 2 x 2 within-subjects design, varying three features of the LLM responses: accuracy of the LLM's answer to the question (correct/incorrect), presence of an explanation (absent/present), and presence of clickable sources (absent/present). Further, we capitalize on the natural inconsistencies that arise in LLM responses to investigate the effects of inconsistencies. We examine the impact of these variables on participants' reliance, accuracy, and other measures, such as confidence, source clicking behavior, time on task, evaluation of LLM responses, and likelihood of asking follow-up questions. \looseness=-1

We find that when either or both an explanation and sources are present, participants report higher confidence in their answer, rate the LLM response higher in terms of the quality of the justification it provides for the answer and the actionability of its response, and are less likely to ask follow-up questions. However, explanations and sources differ in their effects on reliance. Explanations increase reliance on both correct LLM answers and incorrect LLM answers. In contrast, sources increase appropriate reliance on correct LLM answers, although less effectively than explanations, while decreasing overreliance on incorrect LLM answers. Finally, when explanations have inconsistencies, we observe less overreliance on incorrect LLM answers compared to when there are no inconsistencies or when explanations are not provided at all. 
We complement these quantitative findings with qualitative insights and close with a discussion of implications and future research directions for fostering appropriate reliance on LLMs.

Together, our approach and findings offer a number of contributions. (1) Our studies  tackle the timely and critical issue of fostering appropriate reliance on LLMs. Since research on user reliance on LLMs is relatively new, we take a mixed-methods approach, first (via the think-aloud study) identifying features of LLM responses that shape user reliance, and then (via the controlled experiment) isolating and studying the effects of the identified features. (2) Through our two studies, we identify which combinations of features help people achieve appropriate reliance and high task accuracy, providing actionable insights on how to adjust LLM response features. We also contribute a more holistic and nuanced understanding of user reliance on LLMs with insights on people's interpretation of explanations from LLMs, source clicking behavior, and interaction effects between explanations and sources. (3) We provide an in-depth discussion of the implications of our findings, limitations of our work, and future research directions. In particular, we identify providing (accurate and relevant) sources and highlighting inconsistencies and other unreliability cues in LLM responses as promising strategies for fostering appropriate reliance on LLMs. However, such approaches should always be tested with users before deployment. \looseness=-1

\section{Related Work}

\subsection{Appropriate Reliance on AI}

Despite the rapid progress of technology, AI systems still frequently and unexpectedly fail. Without knowing when and how much to rely on a system, a user may experience low-quality interactions or even safety risks in high-stakes settings. 
Prior work has investigated how providing information about an AI system's accuracy \cite{yin2019understanding,He2023Accuracy,Yu2019IUI} and (un)certainty \cite{zhang2020effect,Bansal2021CHI,Green2019,Bucinca2021CSCW,Bussone2015}, explanations of outputs \cite{zhang2020effect,Gonzalez2021DoEH,Bansal2021CHI,Lai2019FAccT,Green2019,Bucinca2021CSCW,Bussone2015}, and onboarding materials \cite{Cai2021,Lai2020Tutorial} impact user reliance, as well as the roles played by human intuition \cite{chen2023understanding}, task complexity \cite{Salimzadeh2023UMAP,Salimzadeh2024CHI}, and other human, AI, and context-related factors \cite{Kim2023Trust}.
However, fostering appropriate reliance on AI remains difficult.
Findings on the effectiveness of proposed methods are mixed, and more research is needed on how reliance is shaped in real-world settings.

While most prior work on AI reliance has been in the context of classical AI models (e.g., specialized classification models), there is a growing body of work looking at reliance on systems based on LLMs or other modern generative AI models \cite{vasconcelos2023generation,spatharioti2023comparing,zhou2024relying,Kim2024FAccT,si2024fact,Lee2024FAccT}. 
For example, several recent studies explored the effect of communicating (un)certainty in LLMs by highlighting uncertain parts of LLM responses \cite{vasconcelos2023generation,spatharioti2023comparing} or inserting natural language expressions of uncertainty \cite{zhou2024relying,Kim2024FAccT}, finding that some but not all types of (un)certainty information help foster appropriate reliance.

Contributing to this line of work, we first take a bottom-up approach to identify the features of LLM responses that impact user reliance in the context of answering objective questions with the assistance of a popular LLM-infused application ChatGPT (\cref{sec:study1}).
In line with findings from prior work~\cite{si2024fact}, we see that reliance is shaped by the content of \textit{explanations} provided by the system, particularly whether or not these explanations contain \textit{inconsistencies}. We also observe that participants seek out \textit{sources} to verify the information provided in responses. We then design a large-scale, pre-registered, controlled experiment to isolate and study the effects of these features (\cref{sec:study2}). We discuss the relevant literature on these features and their impact on AI reliance next.

\subsection{Explanations and Inconsistencies}
\label{sec:llmresponses}

The impact of \emph{explanations} on human understanding and trust of AI systems has been studied extensively within the machine learning and human-computer interaction communities, often under the names explainable AI or interpretable machine learning~\cite{liao2021human,vaughan2021humancentered,arrieta2019explainable,RudinEtAlSurvey2022,Kim2023CHI}. Explanations are often motivated as a way to foster appropriate reliance and trust in AI systems, since in principle they provide clues about whether a system's outputs are reliable. However, empirical studies have shown mixed results, with a large body of work suggesting that providing explanations increases people's tendency to rely on an AI system even when it is incorrect~\cite{zhang2020effect,Bansal2021CHI,Poursabzi-Sangdeh-CHI2021,wang2021explanations}. One potential reason for this is that study participants do not make the effort to deeply engage with the explanations~\cite{kaur2020CHI,buccinca2020proxy,gajos2022people,liao2022designing,Vasconcelos2023CSCW}. That is, instead of encouraging deep, analytical reasoning (System 2 thinking~\cite{Kahneman2003,kahneman2011thinking}), study participants may resort to heuristics, such as the explanation's fluency or superficial cues to expertise~\cite{trout2008}, and defer to the system's response on this basis. People may also be more likely to assume an AI system is trustworthy simply because it provides explanations~\cite{ehsan2021explainable}. Further, some clues of unreliability may be difficult to pick up on without existing domain knowledge~\cite{chen2023understanding}.

Adopting the broad definition of an explanation as an answer to a why question \cite{Lombrozo2012,Wellman2011,bromberger1966why,Fraassen1980}, LLMs often provide explanations by default; when asked a question, LLMs rarely provide the answer alone. For factual questions, they provide details supporting the answer \cite{Lee2024FAccT,lfqa23}, and for math questions, they provide detailed steps to derive the answer \cite{Collins2024PNAS,hendrycks2021measuring}. 
This default behavior is likely due to human preference for verbose responses \cite{chiang2024overreasoning,saito2023verbosity,Zheng2020Verbose}.
Research in psychology has shown that explanations are often sought spontaneously \cite{malle1997behaviors,frazier2009preschoolers}, favored when they are longer, more detailed, or perceived to be more informative \cite{weisberg2015deconstructing,Zemla2017Everyday,bechlivanidis2017concreteness,Aronowitz2020TCS,Liquin2022Satisfaction}, and used to guide subsequent judgments and behaviors \cite{Lombrozo2023Selective,Lombrozo2016}. 
Since LLMs are often fine-tuned on human preference data via approaches such as Reinforcement Learning from Human Feedback (RLHF) \cite{ziegler2019finetuning,Christiano2017RLHF,Ouyang2024RLHF}, such preferences would shape the form of their outputs. We note that the default explanations that LLMs present typically provide evidence to support their answers, but do not necessarily reflect the internal processes by which the LLM arrived at the answer. This distinguishes these explanations from those traditionally studied in the explainable AI literature.

Explanations generated by LLMs are widely known to contain inaccurate information and other flaws \cite{Ji2023Hallucination,shuster2021eval,santhanam2022rome,Dahl2024Law}. We direct readers to recent surveys for comprehensive overviews \cite{huang2023hallucination,wang2023factuality}. In our studies, we found \textit{inconsistencies} in explanations to be an important unreliability cue that shapes participants' reliance. As documented in prior work, inconsistencies can occur within a response; they are sometimes referred to as logical fallacies or self-inconsistency in the NLP community \cite{huang2023reasoning,wang2023selfconsistency}. Inconsistencies can also occur between responses; many studies have demonstrated that LLMs often change their answer to a question when challenged, asked the question in a slightly different way, or re-asked the exact same question \cite{Lee2024FAccT,Elazar2021MeasuringAI,laban2024flipflop}. Such inconsistencies, when noticed, may impact people's evaluation of explanations and reliance on LLMs.

We contribute to this line of work in several ways. We first conduct a qualitative, think-aloud study to understand what features of LLM responses shape people's reliance, and find that reliance is shaped by explanations, inconsistencies in explanations, and sources. We then conduct a larger-scale, pre-registered, controlled experiment to quantitatively examine the effects of these features.
While a previous work by \citet{si2024fact} has studied the effects of LLM-generated explanations and inconsistencies on people's fact-checking performance through a small-scale study (16 participants per condition), our work provides a more holistic picture by studying what (else) might contribute to reliance and how the identified features affect a wider range of variables including people's evaluation of the LLM response's justification quality and actionability and likelihood of asking follow-up questions. 
As for the findings, first, consistent with \citet{si2024fact}, we find that explanations increase people's reliance, including overreliance on incorrect answers, and that inconsistencies in explanations can reduce overreliance. 
Additionally, we find that clickable sources --- which were not studied by \citet{si2024fact} --- increase appropriate reliance on correct answers, while reducing overreliance on incorrect answers, adding empirical knowledge on user reliance on LLMs. 
Lastly, our work also contributes nuanced insights on people's interpretation of LLMs' explanations, source clicking behavior, and interaction effects between explanations and sources.

\subsection{Sources}

The final feature of LLM responses that we study is the presence of \emph{sources}, i.e., clickable links to external material.\footnote{One might consider sources to be a component of an explanation. To simplify the exposition of our results, we treat them as a distinct component of LLM responses throughout this paper.} Sources are increasingly provided by LLM-infused applications, including general-purpose chatbots (e.g., ChatGPT, Gemini) and search engines (e.g., Perplexity AI, Copilot in Bing, SearchGPT). Sources are commonly sought by users, as found in prior work \cite{Kim2024ChatGPT} and supported in our studies. Similar to explanations, however, sources in LLM responses can be flawed in various ways \cite{liu2023evaluating,Alkaissi2023}. For instance, \citet{liu2023evaluating} conducted a human evaluation of popular LLM-infused search engines and found that their responses frequently contain inaccurate sources and unsupported statements. \citet{Alkaissi2023} conducted a case study of ChatGPT in the medical domain and found that it generates fake sources. These issues were observed in our studies as well. Currently there is active research on techniques such as Retrieval Augmented Generation (RAG) \cite{Lewis2020RAG,gao2024rag} to help LLMs provide more accurate information and sources.

It is well known that the presence and quality of sources impact how credible people find given content in other settings \cite{Rieh2007Credibility,Wathen2002Credibility}. However, there has been little work studying how people make use of and rely on sources in the context of LLM-infused applications. On the one hand, the presence of sources might reduce overreliance if people click on the provided links to verify the accuracy of the LLM's response. On the other hand, the presence of sources might increase reliance if people interpret them as signs of credibility and defer to the system without verifying the answers themselves. Indeed, in one study of uncertainty communication in LLM-infused search, participants were found to rarely click on source links~\cite{Kim2024FAccT}. Through a large-scale, pre-registered, controlled experiment (\cref{sec:study2}), we study how the presence of clickable sources impacts people's reliance, task accuracy, and other measures, and how this interacts with the presence of explanations and inconsistencies. In our studies, we use realistic explanations and sources, generated by state-of-the-art LLM-infused applications ChatGPT and Perplexity AI, and provide insights for fostering appropriate reliance on LLMs. \looseness=-1

\begin{figure*}[t!]
\centering
\includegraphics[width=\textwidth]{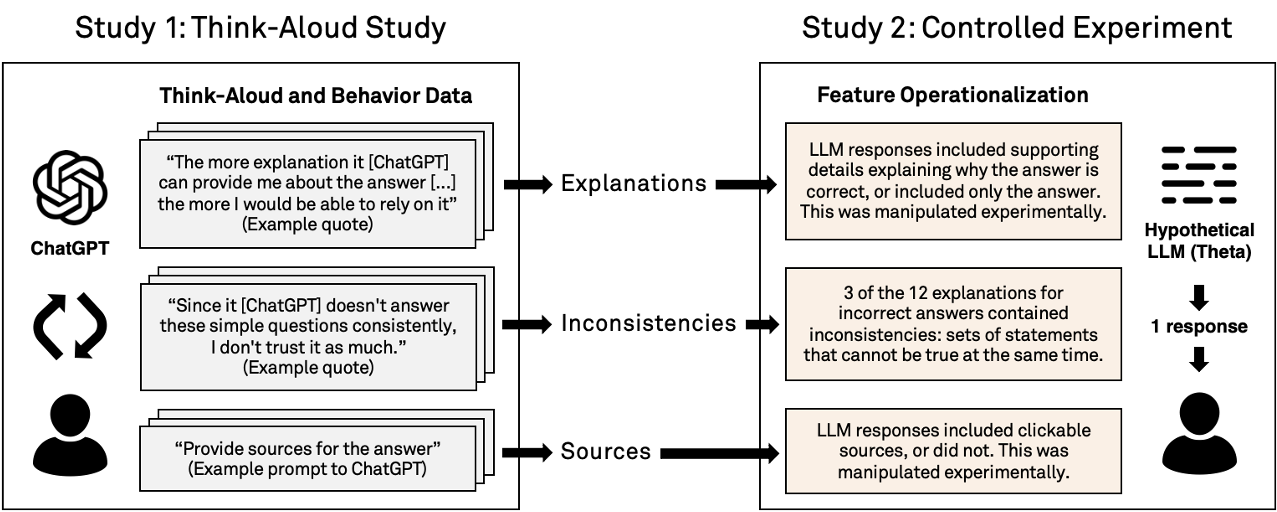}
\caption{\textbf{Overview of our studies.} In Study 1, participants engaged in multi-turn interactions with ChatGPT to arrive at correct answers to objective questions. Based on a thematic analysis of think-aloud and behavioral data, we identified \textit{explanations}, \textit{inconsistencies}, and \textit{sources} as three features of LLM responses likely to influence user reliance. These three features were then investigated in a controlled experiment (Study 2), with features operationalized as indicated in the schematic illustration. Similar to Study 1, participants solved question-answering tasks. However, this time, they had access to one LLM response whose features we experimentally manipulated.}
\label{fig:schematic}
\end{figure*}

\section{Study 1: Think-Aloud Study}
\label{sec:study1}

Towards the goal of identifying features of LLM responses that can help foster appropriate reliance, we first take a bottom-up approach and conduct a think-aloud study in a relatively natural setting. Specifically, we observe how participants solve question-answering tasks with ChatGPT in multi-turn interactions, and explore how they perceive ChatGPT's responses and what helps them arrive at correct answers despite incorrect answers from ChatGPT.

\subsection{Study 1 Methods}

In this section, we describe our study methods, all of which were reviewed and approved by our Institutional Review Board (IRB) prior to conducting the study.

\subsubsection{Procedure}

The study session had two parts.
In Part 1 (Base), participants were introduced to the study and asked to complete three question-answering tasks while thinking aloud. Each task involved determining the correct answer to an objective question using ChatGPT\footnote{We created a research account with a Plus subscription. Participants logged into our account and used ChatGPT-4o --- the latest version at the time (June 2024) --- through the web interface with Browsing allowed and Memory disallowed.} and reporting confidence in their final answer on a 1--7 scale. As in natural settings, participants could exchange as many messages with ChatGPT as they wished. Participants could also check the sources provided in ChatGPT's responses, but were asked not to conduct their own internet search.

Each participant was given three questions: a general domain factual question (e.g., ``Has Paris hosted the Summer Olympics more times than Tokyo?''), a health or legal domain factual question (e.g., ``Is it illegal to collect rainwater in Colorado?''), and a math question (e.g., ``Sue puts one grain of rice on the first square of a Go board and puts double the amount on the next square. How many grains of rice does Sue put on the last square?'').
The factual questions were binary questions. The math questions were not binary, but had one correct numerical answer.
The specific question was randomly selected from a set of questions we created in advance based on examples of real user-LLM interactions \cite{ShareGPTanalysis} and prior work \cite{Shi2023ICML,xie2024adv}.
Before beginning the tasks, we also asked each participant if they knew the answer to any of the questions so that we could switch to a different question if they did, but this did not happen.

In Part 2 (Prompting), we asked participants to complete the same three tasks again, but this time while employing follow-up prompts in their engagement with ChatGPT.
We designed Part 2 to explore whether certain prompts can help participants more appropriately rely on ChatGPT and succeed on the tasks.
Since participants had different levels of familiarity with prompting, we provided examples of prompts they could use, such as asking for a different type of explanation (e.g., ``Explain step by step'' and ``Explain like I'm five''), asking for more information (e.g., ``Provide an explanation with supporting sources'' and ``Explain how confident you are in the answer''), and challenging the previous response (e.g., ``Explain why your answer may be wrong'' and ``I think you are wrong. Try again''). Participants could use whichever and as many prompts as they wished.
As in Part 1, participants reported their final answer and confidence in their final answer at the end of each task.

In between Part 1 and Part 2 and before concluding the study, we asked interview questions about participants' perception of and experience with ChatGPT. Details are in the appendix. \looseness=-1

\subsubsection{Participant recruitment and selection}
\label{sec:study1participants}

To recruit participants, we posted a screening survey on Mastodon, X (previously Twitter), and various mailing lists and Slack workspaces within and outside the first author's institution.
The survey included questions about the respondent's knowledge and use of LLMs.
Based on the survey responses, we selectively enrolled participants to maximize the diversity of the study sample's LLM background.
See below for a summary of participants' knowledge and use of LLMs. 
We manually reassigned two participants to different categories than what they selected in their survey when their survey responses did not line up with their described experience (high to low knowledge for one participant and low to high knowledge for another). 
We refer to individual participants by identifier P\#.
\begin{itemize}[noitemsep,topsep=0pt]
    \item \textit{Low-knowledge}: ``Slightly familiar, I have heard of them or have some idea of what they are'' \interview{6,9,13,15} or ``Moderately familiar, I know what they are and can explain'' \interview{2,3,11,14}.
    \item \textit{High-knowledge}: ``Very familiar, I have technical knowledge of what they are and how they work'' \interview{1,4,8,10,16} or ``Extremely familiar, I consider myself an expert on them'' \interview{5,7,12}.
    \item \textit{Low-use}: ``Never'' use LLMs \interview{5,13,15,16} or use LLMs ``Rarely, about 1--2 times a month'' \interview{4} or use LLMs ``Sometimes, about 3-4 times a month'' \interview{3,6,8}.
    \item \textit{High-use}: Use LLMs ``Always, about once or more a day'' \interview{1,2,7,9,10,11,12,14}.
\end{itemize}

\subsubsection{Conducting and analyzing studies} 

We collected data from 16 participants in June 2024, each over a Zoom video call. The study lasted one hour on average, and participants were paid \$20 for their participation.
All sessions were video recorded and transcribed for data analysis.
We used a mix of quantitative and qualitative methods to analyze the study data.
On the quantitative side, we analyzed the accuracy of participants' answers and their self-reported confidence in their answers measured on a 1--7 scale for each task.
Since each participant solved three tasks, once in Part 1 and again in Part 2, there are 6 accuracy and 6 confidence numbers for each participant.
On the qualitative side, we conducted a thematic analysis~\cite{boyatzis1998transforming,BraunClarke2006} of participants' think-aloud data and their responses to interview questions to identify features of LLM responses that shaped participants' reliance.
The first author performed the initial coding, discussed the categories with other authors, and then refined the coding. \looseness=-1

\subsection{Study 1 Results}
\label{sec:study1task}

We first provide some descriptive statistics about participants' accuracy, over- and underreliance, and confidence across the two parts of the study (\cref{sec:study1quantitative}). 
We then discuss which LLM response features participants reported as influences on their reliance (\cref{sec:study1qualitative}). 
We emphasize that this study was not intended to provide statistically significant results, but to identify features that may help foster appropriate reliance. Given the small sample size, we report the quantitative results only to provide context.

\subsubsection{Accuracy, reliance, and confidence}
\label{sec:study1quantitative}

In Part 1 (Base), we collected data on 48 task instances (16 participants $\times$ 3 tasks). 
For 34 of these instances, ChatGPT gave a correct answer in its first response. (ChatGPT sometimes changed its answer over the course of the interaction, either due to stochasticity or in response to participants' follow-up messages.)
Among these, participants' final answer agreed with ChatGPT's correct answer in 33 instances (average confidence 5.97 on the 1--7 scale) and disagreed in only a single instance (confidence 4.5), indicating that \textbf{underreliance was not prevalent}.
In 13 instances, ChatGPT gave an incorrect answer in its first response. 
Among these, participants' final answer agreed with ChatGPT's incorrect answer in 9 instances (average confidence 6.15) and disagreed in only 4 instances (average confidence 5.61), indicating \textbf{widespread overreliance}.
In a single instance, ChatGPT did not answer the question in its first response, and the participant submitted an incorrect answer with a confidence of 2.

We did not find meaningful differences in participants' accuracy between the 
two parts of the study. That is, \textbf{follow-up prompting did not increase participants' accuracy}, at least based on our small sample of quantitative data.
For 44 out of 47 instances in which the participant completed Part 2 (Prompting) (one participant had to skip a task instance due to lack of time), the participant submitted the same answer in both parts.
In 3 instances, participants submitted an incorrect answer in Part 1 and a correct answer in Part 2.
In 2 of these 3 instances, ChatGPT gave an incorrect answer in Part 1, but gave a correct answer in Part 2. In the other instance, ChatGPT gave incorrect answers in both parts, but the participant arrived at the correct answer in Part 2 after engaging in multiple rounds of interaction with ChatGPT.

Finally, we compared participants' confidence in their answers for the same task between the two parts, finding that it increased in Part 2 in 19 instances, decreased in 8 instances, and stayed the same in 20 instances.
However, \textbf{changes in confidence do not correspond to changes in answers}. As mentioned above, participants changed their answers in only 3 out of 47 instances. In these 3 instances, participants' confidence stayed the same or increased slightly as their answer changed from being incorrect to correct.
Participants' self-described reasons for increased confidence included seeing and checking sources, seeing ChatGPT give the same answer multiple times, and receiving more information in general.
Reasons for decreased confidence included experiencing issues with sources (e.g., links were broken or sources were not reputable) and seeing ChatGPT change answers.

\subsubsection{LLM response features shaping reliance}
\label{sec:study1qualitative}

From a thematic analysis of participants' think-aloud data and responses to interview questions, we found \textbf{explanations}, \textbf{inconsistencies}, and \textbf{sources} to be key features of LLM responses that participants reported as influences on reliance.
First, consistent with our discussion in \cref{sec:intro,sec:llmresponses}, we observed that ChatGPT provided \textbf{explanations} of its answers by default.
Participants found these explanations important for judging the reliability of ChatGPT's answers.
For example, P14 (low-knowledge, high-use) described explanations as \shortquote{very important for having reliability on the answer} and said \shortquote{the more explanation it [ChatGPT] can provide me about the answer [...] the more I would be able to rely on it.}
P11 (high-knowledge, high-use) added that they judge the response by \shortquote{how well ChatGPT explains the answer.}
This participant judged ChatGPT's explanation in one task to be very high quality, noting \shortquote{I would put this on my homework and submit it [...] the quality is very high}.

However, in another task, P11 submitted a different answer from ChatGPT after observing \textbf{inconsistencies}: \shortquote{Since it [ChatGPT] doesn't answer these simple questions consistently, I don't trust it as much.}
Sometimes inconsistencies occurred within a response (e.g., ChatGPT saying Paris hosted the Summer Olympics more times than Tokyo while also saying both have hosted twice). At other times inconsistencies occurred across multiple responses (e.g., ChatGPT changing its answer when asked the same or similar questions, or when challenged).
In either case, \textbf{when participants observed inconsistencies, they often asked follow-up questions and engaged more with the system to resolve the inconsistencies.}
For example, when P8 (high-knowledge, low-use) was considering the question ``Did Tesla debut its first car model before or after Dropbox was founded?'' ChatGPT initially stated that Tesla debuted its first car model in 2008 then later changed the year to 2006. After noticing the inconsistencies, P8 engaged in three more rounds of interaction with ChatGPT to verify individual pieces of information, and arrived at the correct answer.

Finally, participants frequently sought and used \textbf{sources} to determine whether or not to rely on ChatGPT.
More often than not, ChatGPT did not provide sources as part of its responses, even though participants were using the latest version at the time of the study (4o) with browsing capabilities. Participants had to explicitly ask for them using prompts like ``Provide sources for the answer.''
Participants rarely did this in Part 1, and as such, sources were provided in only 17 out of 48 instances. However, in Part 2, participants asked for sources more often after seeing prompt examples and were provided sources in 30 instances.
\textbf{When participants checked sources, they were often able to avoid overreliance on ChatGPT.}
For example, out of 11 instances in which participants submitted correct answers despite incorrect answers from ChatGPT (both parts combined), 7 were instances in which participants checked sources. (In the other 4 instances, sources were not provided, but participants were able to submit correct answers through other strategies, such as repeatedly asking ChatGPT about a piece of information.)
For example, when P2 (low-knowledge, high-use) was solving the question ``Sue puts one grain of rice on the first square of a Go board and puts double the amount on the next square. How many grains of rice does Sue put on the last square?'' ChatGPT built on an incorrect assumption about the size of a Go board and gave an incorrect answer. P2 initially judged it as correct, but after checking sources, realized ChatGPT's error and was able to submit a correct answer.

As discussed in \cref{sec:study1quantitative}, sources also influenced participants' confidence in their answers. \textbf{The presence of sources increased confidence in general, except when there were issues with sources.}
For example, P1 (high-knowledge, high-use) said their confidence increased in Part 2 for one task when they received sources and were able to verify information in ChatGPT's responses. But they said their confidence decreased for another task when some of the source links did not open or did not contain relevant information, highlighting the importance of source \textit{quality} in addition to \textit{presence}.
Finally, we emphasize that \textbf{checking sources did not always eliminate overreliance}.
Out of 30 instances in which participants checked sources (both parts combined), in 4 instances, participants' final answer still agreed with ChatGPT's incorrect answer, which is a sign of overreliance.

\begin{figure*}[t!]
\centering
\includegraphics[width=\textwidth]{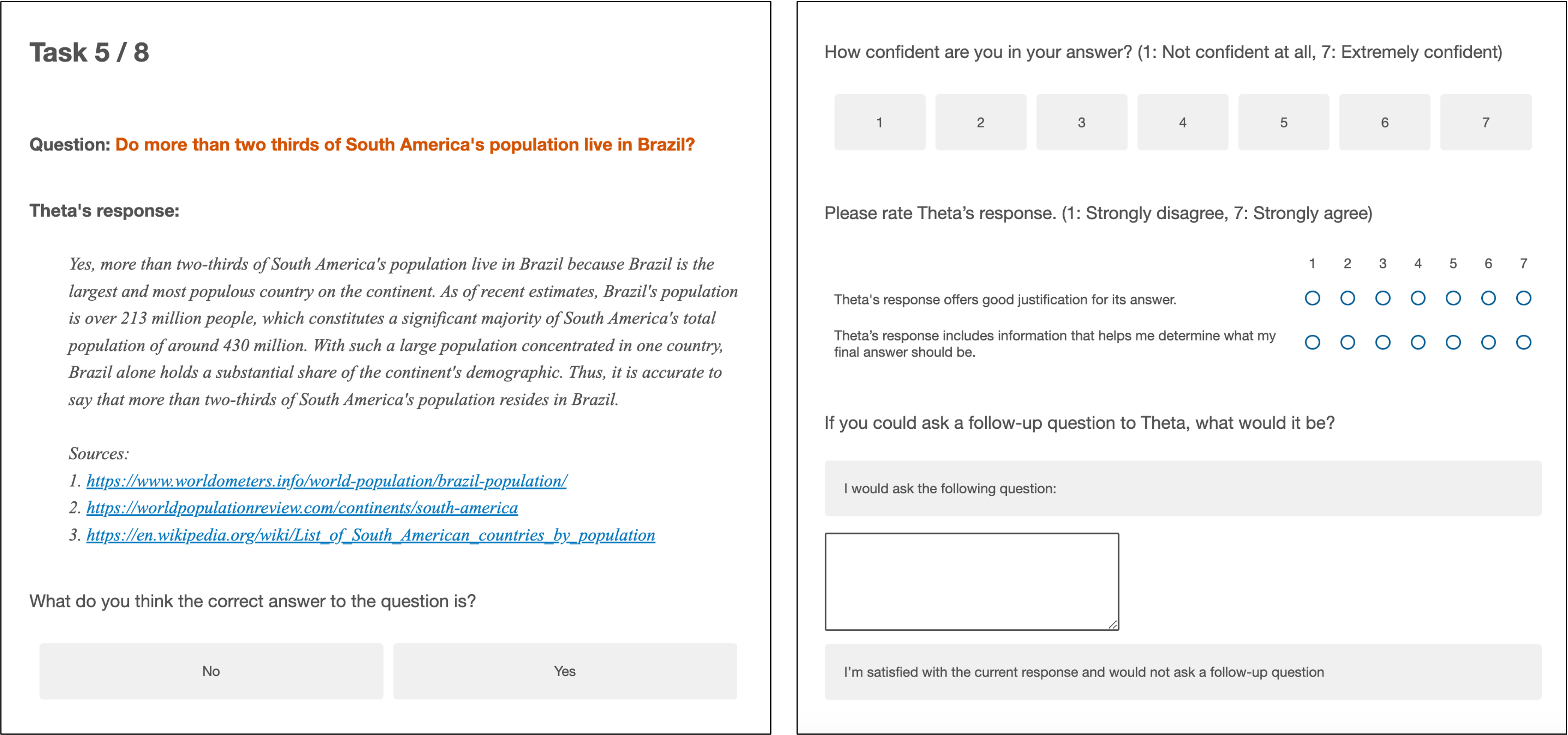}
\caption{\textbf{Screenshots of Study 2's experimental task.} Here the LLM response provides an incorrect answer, includes sources, and includes an explanation (with inconsistencies). See \cref{fig:types} for responses with a correct answer for the same task question.}
\label{fig:task}
\end{figure*}

\section{Study 2: Large-scale, Pre-registered, Controlled Experiment}
\label{sec:study2}

Based on the insights from Study 1, we designed a large-scale, pre-registered, controlled experiment to study the effects of different features of LLM responses on people's reliance, task accuracy, and other measures including confidence, source clicking behavior, time on task, evaluation of LLM responses, and asking of follow-up questions.
The goal of the study was to test whether the findings from Study 1 apply at scale and identify which combinations of features help people achieve appropriate reliance and high task accuracy. \looseness=-1

\subsection{Study 2 Methods}

In this section, we describe our study methods. Before collecting data, we obtained IRB approval and pre-registered our experimental design, analysis plan, and data collection procedures.\footnote{Our pre-registration is viewable at \url{https://aspredicted.org/bg22-yfw7.pdf}.}

\subsubsection{Procedure}

\begin{figure*}[t!]
\centering
\includegraphics[width=\textwidth]{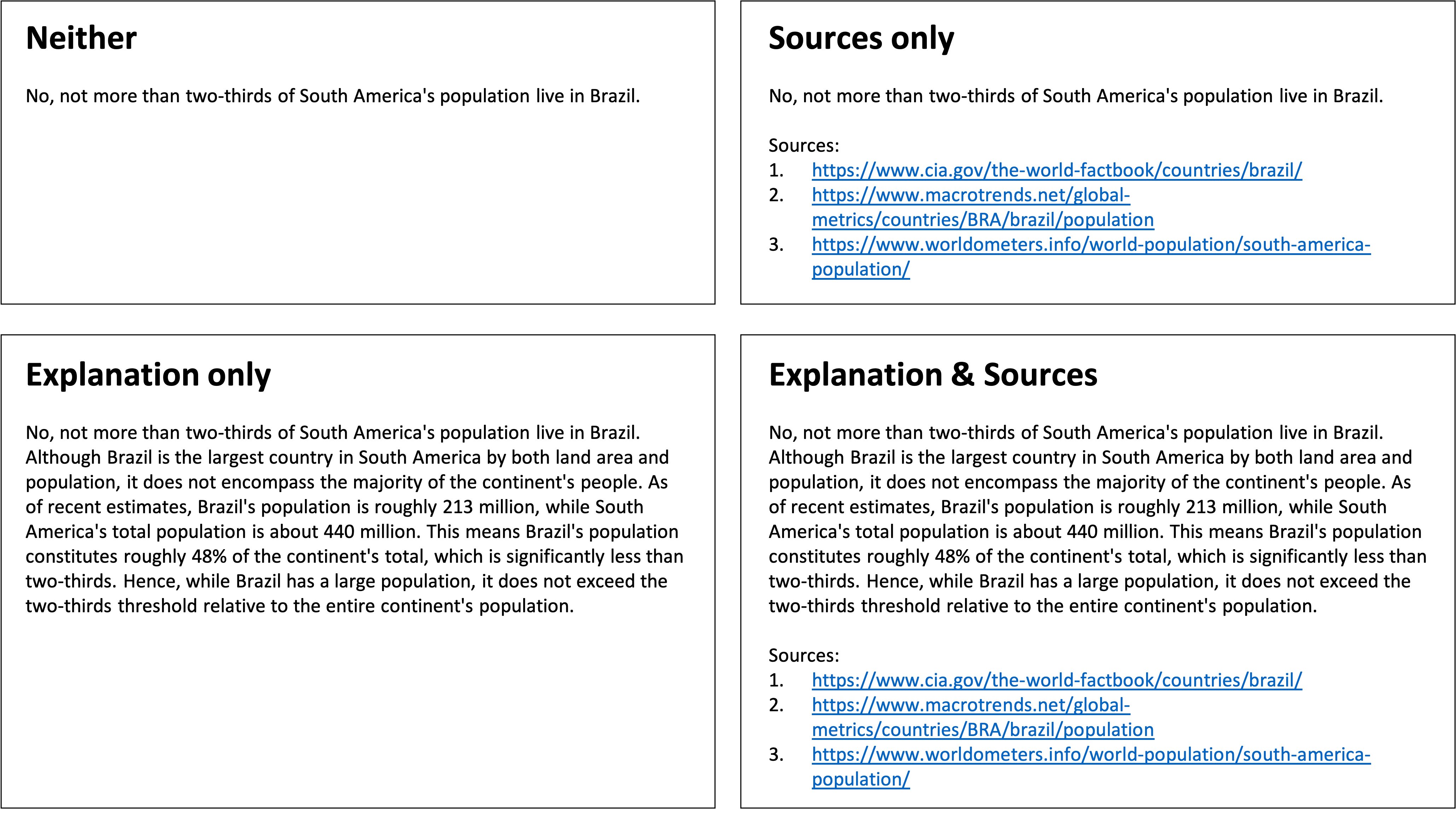}
\caption{\textbf{Types of LLM responses used in Study 2.} We vary three variables in the LLM responses: accuracy of the LLM's answer to the question (correct/incorrect), presence of an explanation (absent/present), and presence of clickable sources (absent/present). In total there are 8 types of responses. Here we show 4 types of responses with a correct answer to the question: ``Do more than two thirds of South America's population live in Brazil?'' See \cref{fig:task} for a response with an incorrect answer.}
\label{fig:types}
\end{figure*}

We designed a within-subjects experiment in which participants completed a set of question-answering tasks with LLM responses.
Each task involved determining the correct answer to a binary factual question with access to a response from a hypothetical LLM named ``Theta'' (hereafter we occasionally refer to it as ``the LLM'').
See \cref{fig:task} for an example.
Our experiment had a 2 x 2 x 2 design where we varied three variables in Theta's responses: accuracy of Theta's answer to the question (correct/incorrect), presence of an explanation (absent/present), and presence of clickable sources (absent/present).
In total, there were 8 types of responses. 
Participants completed 8 tasks in the experiment and saw one of each type.
This makes Theta's accuracy 50\%, but participants were not given this information: participants did not receive feedback on whether their answer or Theta's answer was correct after solving a task.
See \cref{fig:types} for examples of different types of responses.

The experiment had three parts.
In the first part, participants were introduced to the study and to Theta.
Theta was described as an LLM-based AI system prototype that uses similar technology to OpenAI's ChatGPT, is connected to the internet, and can answer a wide range of questions.
In the second part, participants answered a total of eight questions.
For each question, participants were provided with a response from Theta and were asked to submit their answer, report their confidence in their answer, and rate Theta's response.
They were told that they could click on source links in Theta's responses, but asked not to conduct their own internet search.
Participants could also optionally write a follow-up question, but they did not see Theta's response to it. We made this choice to fully control the number and content of responses, while being able to collect data on when and what types of follow-up questions participants ask.
We acknowledge that showing one controlled response instead of allowing free-form interaction has limitations (see Section~\ref{sec:limitations}). However, we adopt this method from prior work studying LLMs~\cite{Kim2024FAccT,Lee2024FAccT,si2024fact} as a valid approach for capturing user perceptions and behaviors around LLM responses with the advantage of controlling unwanted noise from free-form interactions (for instance, LLMs making different mistakes across participants in follow-up interactions).

We randomized the order in which questions were presented, as well as the assignment of the 8 response types to the questions.
In the final part, participants filled out an exit questionnaire about their experience with and perception of Theta, their background on LLMs, and basic demographic information.
Lastly, participants were debriefed and reminded that some of the responses they saw may have contained inaccurate information.

\subsubsection{Dependent Variables}
\label{sec:dvs}

We formed a set of dependent variables (DVs) using a mix of behavioral and self-reported measures to capture participants' reliance and accuracy, as well as related behaviors and judgments.
First, we measured the \emph{agreement} between a participant's answer and that of Theta; this is a commonly used behavioral measure of reliance \cite{yin2019understanding,zhang2020effect,Lai2019FAccT,Bucinca2021CSCW,Cao2022CSCW,Liu2021CSCW,Lu2021CHI,Mohseni2020}.
Second, we measured the \emph{accuracy} of a participant's answer to assess the task outcome. These are our main two DVs.
To complement them, we also examined participants' \emph{confidence} and \emph{source clicking behavior} as indirect measures of reliance, as well as \emph{time on task}, since efficiency is also an important aspect of task outcome.
These complementary measures have also been commonly studied in prior work~\cite{Poursabzi-Sangdeh-CHI2021,Cao2022CSCW,Kim2022HIVE,CHONG2022,Lu2021CHI,vasconcelos2023generation,Kim2024FAccT}.

Additionally, we had participants evaluate the individual LLM responses.
First, we had participants evaluate the \emph{justification quality} of a response, i.e., whether it offers a good justification for its answer. 
Based on prior work in psychology, we expected this to be correlated with reliance and confidence \cite{Lombrozo2016,douven2018best}, as well as whether participants ask follow-up questions \cite{Liquin2022Satisfaction,frazier2009preschoolers}.
Second, we had participants evaluate the \emph{actionability} of a response, as incorrect responses or responses with low justification quality can still be useful if they are actionable; recall that in Study 1, we observed that participants often treated an LLM response as a starting point for determining what action to take next to arrive at the correct answer. 
Finally, we measured whether participants wrote a follow-up question they would like to ask to Theta. This is in part a proxy for satisfaction: prior work in psychology has found that children are less likely to re-ask a question when they are satisfied with an initial response \cite{Kurkul2018Followup,Frazier2016Satisfying,Mills2017Children}. On the other hand, greater satisfaction with a response can increase curiosity about related content \cite{Liquin2022Satisfaction}.

Formally, we measured the following DVs based on participants' observed behavior:
\begin{itemize}
    \item \var{Agreement}: TRUE if the participant's final answer is the same as Theta's answer; FALSE otherwise.
    \item \var{Accuracy}: TRUE if the participant's final answer is correct; FALSE otherwise.
    \item \var{SourceClick}: TRUE if the participant clicked on one or more sources; FALSE otherwise.
    \item \var{Time}: Number of minutes from when the participant saw the question to when they clicked next.
\end{itemize}

We additionally measured the following DVs based on participants' self-reported ratings or selections:
\begin{itemize}
    \item \var{Confidence}: Rating on the question ``How confident are you in your answer?'' on a 7-point scale. 
    \item \var{JustificationQuality}: Rating on the statement ``Theta's response offers good justification for its answer'' on a 7-point scale. 
    \item \var{Actionability}: Rating on the statement ``Theta's response includes information that helps me determine what my final answer should be'' on a 7-point scale. 
    \item \var{Followup}: TRUE if the participant wrote a follow-up question they would like to ask instead of selecting ``I'm satisfied with the current response and would not ask a follow-up question.''
\end{itemize}

All DVs were measured once for each of the 8 tasks.
See \cref{fig:task} for screenshots of an example task.

\subsubsection{Analysis}
\label{sec:study2analysis}

We hypothesized that the three features of LLM responses that we manipulated --- the accuracy of the answer, the presence of sources, and the presence of an explanation --- would affect each of the DVs.
To examine this hypothesis, we used a mixed-effects regression model (logistic or linear depending on the data type), where each participant has a unique ID and each task question has a unique ID.
Specifically, for each DV except \var{SourceClick}, we fit the model 
\texttt{DV $\sim$ AI\_Correct * AI\_Sources * AI\_Explanation + (1|participant) + (1|question)}. 
For \var{SourceClick}, we fit the model \texttt{DV $\sim$ AI\_Correct * AI\_Explanation + (1|participant) + (1|question)} only looking at data points for which participants were provided with sources.
\texttt{AI\_Correct}, \texttt{AI\_Sources}, and \texttt{AI\_Explanation} are binary variables with \texttt{Correct Answer}, \texttt{No Sources}, and \texttt{No Explanation} as the reference levels.

We complemented the main analysis with several additional analyses.
First, we conducted two pre-registered analyses 
exploring how participants reacted to inconsistencies in explanations (\cref{sec:inconsistencies}) and how participants' source clicking behavior relates to other DVs (\cref{sec:sourceclick}).
Analysis details and results are presented in the respective sections.
Second, we conducted a thematic analysis~\cite{boyatzis1998transforming,BraunClarke2006} of participants' free-form answers in the exit questionnaire.
The results are presented in \cref{sec:study2results} alongside the quantitative results from the main analysis.

\subsubsection{Materials}
\label{sec:study2materials}

To simulate a realistic LLM usage scenario of users seeking answers to questions they don't know the answer to, we selected task questions according to the following criteria: (1) most lay people should not know the answer off the top of their head so that they will likely engage with the LLM response and (2) the answer can be objectively and automatically assessed.
To satisfy the criteria, we first created 32 binary factual questions based on facts from the books \textit{Weird But True Human Body}~\cite{WeirdButTrueHumanBody} and \textit{Weird But True World 2024}~\cite{WeirdButTrueWorld2024} by National Geographic Kids. We then ran a short pilot study ($N = 50$) in which we asked participants to answer the 32 questions based on their knowledge and without consulting external sources. This allowed us to assess how commonly known the answers to the questions are in our sample. 
We selected questions with less than 50\% accuracy (i.e., worse than random guessing) as our final set of task questions (12 in total) to satisfy our first selection criterion.
However, we acknowledge that focusing on difficult questions may affect the generalizability of our results. See \cref{fig:task,fig:types} for an example question and the appendix for the full set. \looseness=-1

To create LLM responses that are realistic and reflect the state-of-the-art, we used ChatGPT-4o with a Plus subscription and with Browsing allowed, Memory disallowed, and a new chat for each prompt. 
Initially, we inputted the selected task questions to ChatGPT without any system prompts. Consistent with prior work~\cite{Lee2024FAccT}, we observed that ChatGPT's responses generally follow the same structure: answer to the question (e.g., yes or no) followed by an explanation (supporting details). However, the responses greatly varied in form (e.g., the number of paragraphs and the use of bulleted or numbered lists) and length (ranging from 48 to 213 words). To reduce this variability, we used the system prompt ``Provide a one paragraph response not exceeding 180 words'' following the choices in prior work~\cite{Lee2024FAccT}.

For each task question, we first created a pair of responses with explanations, one with a correct answer and one with an incorrect answer. 
To do so, we used the prompts ``Why is [correct answer/incorrect answer] the correct answer to the question: [Task question]?''
We note that the obtained responses were similar in nature to responses obtained by just asking the task question.
We phrased the prompts this way to reduce any structural differences between responses for correct and incorrect answers.
We made minor edits to improve readability and ensure all responses had the same structure (i.e., answer to the question followed by an explanation). We did not make substantive edits to the content.
To create responses without explanations, we removed the explanation component from these responses.

To create responses with sources, we sent the same prompts to Perplexity AI, one of the most popular LLM-infused search engines, with a Plus subscription and with GPT-4o as the AI model. This is because none of the responses from ChatGPT-4o included sources, even with Browsing allowed.
Perplexity AI's responses included 5 to 10 sources. 
When we analyzed the sources, we found that all sources are real, relevant to the task question, and tended to provide accurate information, although we could not fact-check every single statement in these sources.
To not overwhelm participants, we randomly selected 3 sources and appended them to the responses with/without explanations to create responses with sources.
All responses from ChatGPT-4o and Perplexity AI were retrieved between July 29 and August 1 of 2024 using the latest version at the time. \looseness=-1

After creating different types of LLM responses, we went through the responses again and coded the presence of \textit{inconsistencies}, i.e., sets of statements that cannot be true at the same time~\cite{Hurley2000}, which we found to be an important unreliability cue in Study 1.\footnote{To code inconsistencies, we carefully read each LLM response and reasoned about every pair of statements (whether they can be true at the same time). This was doable because the responses are relatively short (less than 180 words) and do not require specialized knowledge to understand. For the same reasons, we expect most people to be able to notice these inconsistencies. We note that this may not always be the case. The presence of inconsistencies may have little to no effect if they are hard to detect, for example, because the LLM response is long, complex, and requires specialized knowledge to understand that two statements cannot be true at the same time.}
We found that 3 out of 12 responses with explanations for an incorrect answer contained inconsistencies:
(1) For the task question ``Do more than two thirds of South America's population live in Brazil?'' the incorrect response (see \cref{fig:task}) says ``yes'' but later states Brazil's population as around 213 million and South America's total population as around 430 million. (2) For ``Which body part has a higher percentage of water, lungs or skin?'' the incorrect response says ``skin'' but later states skin has 64\% and lungs have 83\% water percentage. (3) For ``Do all mammals except platypus give birth to live young?'' the incorrect response says ``yes'' but later states echidnas as another mammal species that does not give birth to live young.
In contrast, none of the 12 responses with an explanation for a correct answer contained inconsistencies.
While the presence of inconsistencies is not something we control for or manipulate, we coded it to study its effects on reliance and other measures.
See \cref{fig:types} for examples of different types of responses and the appendix for more information on the study materials.

\subsubsection{Participants}

We aimed to collect a minimum of 300 responses post-exclusions.
This number was determined based on a power analysis on pilot data using the simR package in R \cite{simR}.
We conducted data collection using Qualtrics and Prolific in August 2024.
Specifically, we collected responses from 320 U.S.-based adults on Prolific who had completed at least 100 prior tasks with a 95\% or higher approval rating. 
We excluded 12 responses (3.75\%) based on three pre-registered exclusion criteria (3 for response time under 5 minutes, 9 for less than 80\% accuracy on the post-task attention check, and 1 for off-topic free-form response; 1 response was caught on multiple criteria).
Our final sample consists of 308 responses.
Regardless of inclusion or exclusion in the final sample for analysis, we paid all participants \$3.75.
The median study duration was 15.3 minutes, so on average, participants were paid \$14.70 per hour.
See the appendix for more information about participants. \looseness=-1

\subsection{Study 2 Results: Main Analysis}
\label{sec:study2results}

We begin with the main analysis results.
We report the raw data means ($M$) and standard deviations in \cref{tab:main} and the regression results ($\beta, SE, p$) in the text.
We use \textit{significance} to refer to statistical significance at the level of $p < .05$. 
Recall that we fit mixed-effects regression models with three variables and all possible interactions (see \cref{sec:study2analysis} for details).
We did not find a significant three-way interaction for any DVs.
Given our interest in the effects of explanation and sources in LLM responses, we report significant main effects and two-way interactions in the following order: main effects of explanation and interactions with LLM accuracy (\cref{sec:mainexplanation}), main effects of sources and interactions with LLM accuracy (\cref{sec:mainsource}), interactions between explanation and sources (\cref{sec:explsource}), and additional effects of LLM accuracy (\cref{sec:mainaccuracy}).

\begin{table*}[t]
\caption{\textbf{Study 2 main results.} We report the raw data means (and standard deviations) for the eight types of LLM responses: \{Correct \cmark, Incorrect \xmark \, answer\} $\times$ \{Neither, Explanation only, Sources only, Explanation \& Sources\}. See \cref{sec:study2results} for details.}
\label{tab:main}
\begin{tabular}{cccccc}
\hline \\ [-2ex]
DV & Answer & Neither & Explanation only & Sources only & Explanation \& Sources \\ [0.5ex] \hline
\\ [-1.5ex]
\multirow{2}{*}{\var{Agreement} (\%)} & \cmark & 67.2\% (47.0\%) & 78.2\% (41.3\%) & 73.4\% (44.3\%) & 79.9\% (40.2\%) \\ 
 & \xmark & 78.2\% (41.3\%) & 82.8\% (37.8\%) & 68.2\% (46.7\%) & 76.9\% (42.2\%) \\ [2ex]
 
\multirow{2}{*}{\var{Accuracy} (\%)} & \cmark & 67.2\% (47.0\%) & 78.2\% (41.3\%) & 73.4\% (44.3\%) & 79.9\% (40.2\%) \\
 & \xmark & 21.8\% (41.3\%) & 17.2\% (37.8\%) & 31.8\% (46.7\%) & 23.1\% (42.2\%) \\ [2ex]
 
\multirow{2}{*}{\var{Confidence} (1-7)} & \cmark & 4.55 (1.68) & 5.26 (1.47) & 5.50 (1.58) & 5.83 (1.28) \\
 & \xmark & 4.92 (1.73) & 5.47 (1.44) & 5.43 (1.51) & 5.61 (1.28) \\ [2ex]
 
\multirow{2}{*}{\var{SourceClick} (\%)} & \cmark & - & - & 28.2\% (45.1\%) & 25.0\% (43.4\%) \\ 
 & \xmark & - & - & 27.9\% (44.9\%) & 22.1\% (41.5\%) \\ [2ex]
 
\multirow{2}{*}{\var{Time} (min)} & \cmark & 1.05 (1.23) & 1.08 (0.98) & 1.24 (1.04) & 1.30 (1.14) \\
 & \xmark & 0.89 (0.85) & 1.05 (0.94) & 1.39 (1.16) & 1.44 (1.30) \\ [2ex]

\var{Justification} & \cmark & 2.58 (1.89) & 5.52 (1.48) & 4.45 (2.13) & 5.99 (1.32) \\
\var{Quality} (1-7) & \xmark & 2.81 (2.09) & 5.51 (1.62) & 3.90 (2.17) & 5.44 (1.75) \\ [2ex]

\multirow{2}{*}{\var{Actionability} (1-7)} & \cmark & 2.56 (1.94) & 5.14 (1.74) & 4.90 (2.06) & 6.13 (1.19) \\
 & \xmark & 2.91 (2.09) & 5.32 (1.72) & 4.59 (2.09) & 5.62 (1.63) \\ [2ex]
 
\multirow{2}{*}{\var{FollowUp} (\%)} & \cmark & 71.4\% (45.2\%) & 28.2\% (45.1\%) & 34.4\% (47.6\%) & 12.7\% (33.3\%) \\
 & \xmark & 65.3\% (47.7\%) & 27.6\% (44.8\%) & 47.1\% (50.0\%) & 25.6\% (43.7\%) \\ [1ex] \hline
\end{tabular}
\end{table*}

\subsubsection{Main effects of explanation and interactions with LLM accuracy}
\label{sec:mainexplanation}

We find a significant main effect of explanation on most DVs (all except \var{SourceClick} and \var{Time}).
Specifically, provided that the LLM answer is correct and there are no sources, providing an explanation leads to higher participant agreement with the LLM answer ($M = 78.2\%$ vs. $67.2\%$, $\beta = .60, SE = .19, p = .002$), accuracy ($M = 78.2\%$ vs. $67.2\%$, $\beta = .65, SE = .19, p < .001$), confidence in the final answer ($M = 5.26$ vs. $4.55$, $\beta = .74, SE = .10, p < .001$), rating of the LLM response's justification quality ($M = 5.52$ vs. $2.58$, $\beta = 2.94, SE = .13, p < .001$), and rating of its actionability ($M = 5.14$ vs. $2.56$, $\beta = 2.59, SE = .13, p < .001$).
On the other hand, the likelihood of asking a follow-up question is lower when an explanation is provided ($M = 28.2\%$ vs. $71.4\%$, $\beta = -2.38, SE = .21, p < .001$).

For participants' accuracy, however, we find a significant interaction between the presence of an explanation and the accuracy of the LLM answer ($\beta = -1.00, SE = .28, p < .001$).
In the absence of sources, when the LLM answer is correct, participants' accuracy is higher when an explanation is provided ($M = 78.2\%$ vs. $67.2\%$).
In contrast, when the LLM answer is incorrect, accuracy is lower when an explanation is provided ($M = 17.2\%$ vs. $21.8\%$). 
That is, in both cases, participants submitted the same answer as the LLM's more often when an explanation was provided.

We find support for these findings in the qualitative data as well.
In their free-form answers in the exit questionnaire, 28 participants wrote that they submitted a different answer from the LLM's answer when there was no explanation.
As put by one participant, \shortquote{One sentence answers felt incomplete and did not explain how Theta arrived at its conclusion.}
Another wrote the absence of explanation \shortquote{made the [LLM's] answer too hard to trust.}

In summary, we find that \textbf{explanations tend to increase reliance, both appropriate reliance on correct answers and overreliance on incorrect answers}.
Explanations also tend to increase participants' confidence in their answer and evaluation of the LLM response, and decrease their likelihood of asking a follow-up question.
Intuitively, this suggests participants viewed LLM responses with explanations as more satisfying and reliable, regardless of their accuracy.
These findings are consistent with prior research \cite{Bansal2021CHI,zhang2020effect,Poursabzi-Sangdeh-CHI2021,wang2021explanations,Fok2024Verifiability,Pafla2024CHI,si2024fact} and suggest explanations from state-of-the-art LLMs can also lead to overreliance and have unintended negative consequences.

\begin{figure*}[t!]
    \centering
    \begin{subfigure}[t]{0.55\textwidth}
        \centering
        \includegraphics[height=1.6in]{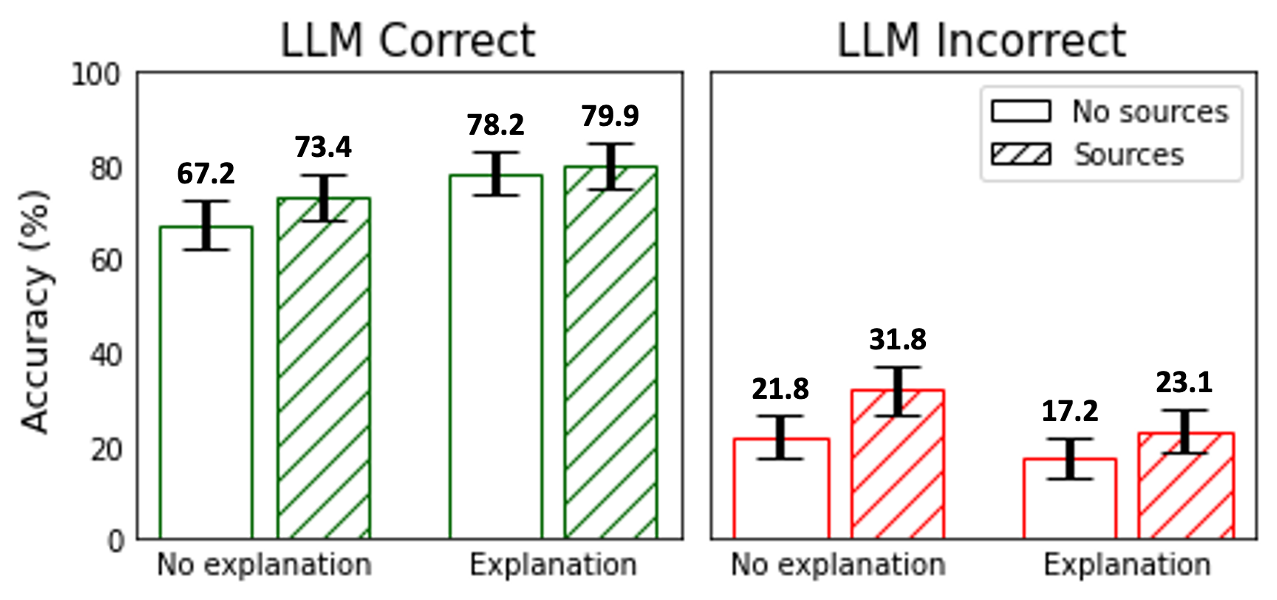}
        \caption{Effect of explanation, sources, and LLM accuracy}
        \label{fig:accuracy_main}
    \end{subfigure}%
    \hfill
    \begin{subfigure}[t]{0.37\textwidth}
        \centering
        \includegraphics[height=1.6in]{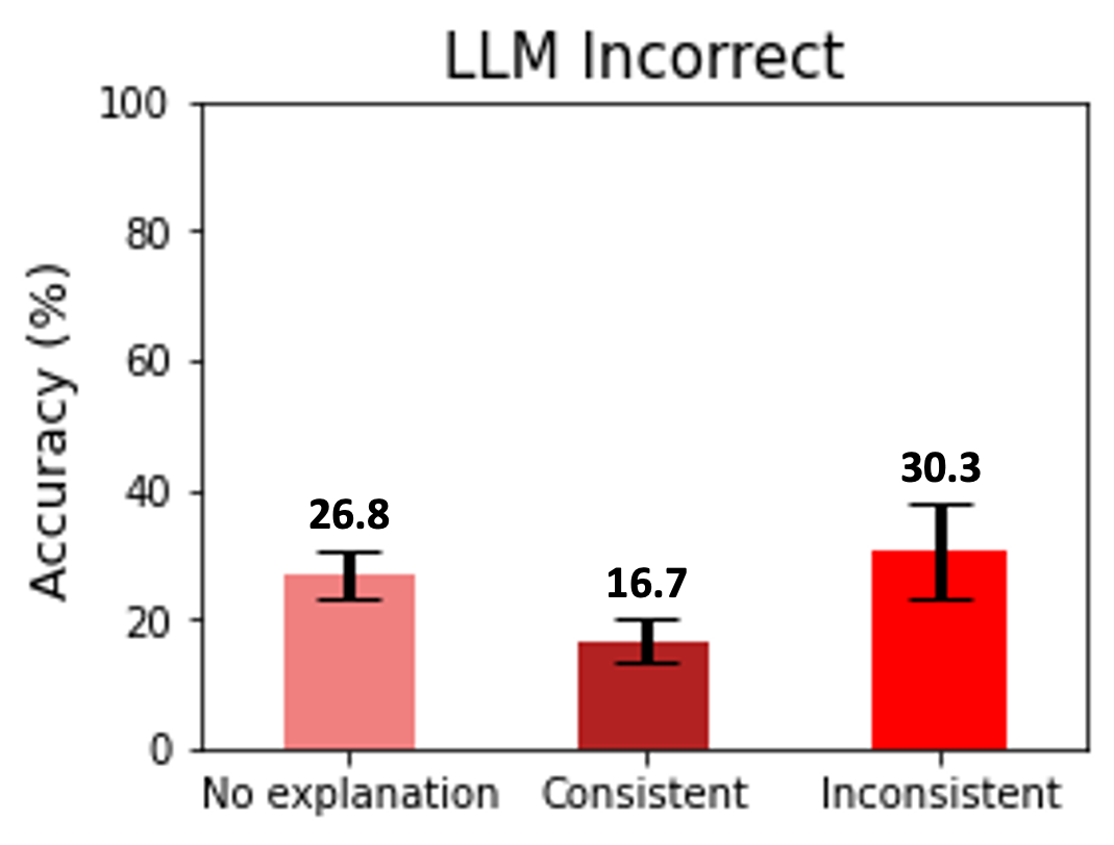}
        \caption{Effect of inconsistencies}
        \label{fig:accuracy_inconsistent}
    \end{subfigure}
    \caption{\textbf{Summary of participants' accuracy in Study 2.} We plot the raw data means and 95\% confidence intervals for participants' accuracy when provided with different types of LLM responses. When the LLM's answer is correct, participants' accuracy is highest when the LLM response includes an explanation and sources (\cref{fig:accuracy_main} left). When the LLM's answer is incorrect, participants' accuracy is highest when the LLM response includes sources but not an explanation (\cref{fig:accuracy_main} right). When the LLM response includes an explanation for an incorrect answer, participants' accuracy is higher when the explanation is inconsistent (\cref{fig:accuracy_inconsistent}).}
\end{figure*}

\subsubsection{Main effects of sources and interactions with LLM accuracy}
\label{sec:mainsource}

We find a significant main effect of sources on the time spent on the task, as well as on all self-reported variables.
That is, when the LLM answer is correct and there is no explanation, providing sources leads to higher participant time on task ($M = 1.24$ min vs. $1.05$ min, $\beta = .17, SE = .07, p = .027$), confidence in the final answer ($M = 5.50$ vs. $4.55$, $\beta = .96, SE = .10 p < .001$), rating of the LLM response's justification quality ($M = 4.45$ vs. $2.58$, $\beta = 1.88, SE = .13, p < .001$), and rating of its actionability ($M = 4.90$ vs. $2.56$, $\beta = 2.34, SE = .13, p < .001$).
In contrast, the likelihood of asking a follow-up question is lower when sources are provided ($M = 34.4\%$ vs. $71.4\%$, $\beta = -2.04, SE = .20, p < .001$).

However, we find a significant interaction between the presence of sources and LLM accuracy on many DVs.
Beginning with agreement ($\beta = -.83, SE = .27, p = .002$),
provided that there is no explanation, when the LLM answer is correct, agreement is higher when sources are provided ($M = 73.4\%$ vs. $67.2\%$).
But when the LLM answer is incorrect, agreement is lower when sources are provided ($M = 68.2\%$ vs. $78.2\%$).
These results suggest that \textbf{sources tend to increase appropriate reliance on correct answers and reduce overreliance on incorrect answers.}

Significant interactions are also found for all self-reported variables: 
\var{Confidence} ($\beta = -.45, SE = .14, p = .002$), \var{JustificationQuality} ($\beta = -.79, SE =.19, p < .001$), \var{Actionability} ($\beta = -.65, SE = .19, p < .001$), and \var{Followup} ($\beta = 1.05, SE = .27, p < .001$).
Provided that there is no explanation and the LLM answer is correct, providing sources increases \var{Confidence}, \var{JustificationQuality}, and \var{Actionability} while decreasing \var{Followup}.
When the LLM answer is incorrect, these effects of sources are all attenuated.
The fact that sources have different effect sizes for correct and incorrect LLM answers provides further (if indirect) support for the idea that sources can help foster appropriate reliance. \looseness=-1

The final significant interaction between sources and LLM accuracy is found for time on task ($\beta = .33, SE = .11, p = .002$).
Provided that there is no explanation, when the LLM answer is correct, time on task is higher when there are sources ($M = 1.24$ min vs. $1.05$ min). But when the LLM answer is incorrect, this effect of sources on time on task is magnified ($M = 1.39$ min vs. $.89$ min).
A possible reason for this result is that when the LLM answer is incorrect, in some instances participants may have found conflicting information between the LLM response and the sources and spent more time resolving the conflict and completing the task.
For example, 45 participants wrote in their free-form responses that they submitted a different answer from the LLM's answer when it conflicted with the information in the sources, e.g., \shortquote{I trusted the information in the links more than I trusted Theta's answer. Therefore, if the information in the links differed, I submitted a final answer that was different from Theta's.} \looseness=-1

Participants also wrote that the mere presence of sources tended to increase the credibility of the LLM response (e.g., \shortquote{If Theta supplied sources for its answers, I felt the answers were more credible}), while the absence of sources had the opposite effect (e.g., \shortquote{Not having any links provided with [Theta's] answer was a red flag to me to think something is wrong or can't be found}).
29 participants explicitly stated that they submitted a different answer from the LLM's answer when there were no sources in the LLM response.
Additionally, several participants wrote about how they were forced to rely on their intuition when there were no sources, e.g., \shortquote{Without being able to verify info, my gut was my best answer.}
They expressed frustration about this and said they would prefer to have sources since it is \shortquote{easiest to agree or disagree when the AI cited its sources.}

\subsubsection{Interactions between explanation and sources}
\label{sec:explsource}

In addition to the main effects of explanation and sources and their respective interactions with LLM accuracy, we find a significant interaction between explanation and sources for all self-reported variables: \var{Confidence} ($\beta = -.42, SE = .14, p = .004$), \var{JustificationQuality} ($\beta = -1.41, SE = .19, p < .001$), \var{Actionability} ($\beta = -1.36, SE = .19, p < .001$), and \var{Followup} ($\beta = .81, SE = .31, p < .001$).
Provided that the LLM answer is correct, when there are no sources, providing explanations increases \var{Confidence} ($M = 5.26$ vs. $4.55$), \var{JustificationQuality} ($M = 5.52$ vs. $2.58$), and \var{Actionability} ($M = 5.14$ vs. $2.56$), while decreasing \var{Followup} ($M = 28.2\%$ vs. $71.4\%$).
When there are sources, however, providing explanations still increases \var{Confidence} ($M = 5.83$ vs. $5.50$), \var{JustificationQuality} ($M = 5.99$ vs. $4.45$), and \var{Actionability} ($M = 6.13$ vs. $4.90$), while decreasing \var{Followup} ($M = 12.7\%$ vs. $34.4\%$), but all to a lesser extent than when there are no sources.
In short, including both explanation and sources achieves the biggest effects in these measures, though their joint effects are subadditive, i.e., less than the sum of the individual effects.

\begin{figure*}[t!]
\centering
\includegraphics[width=\textwidth]{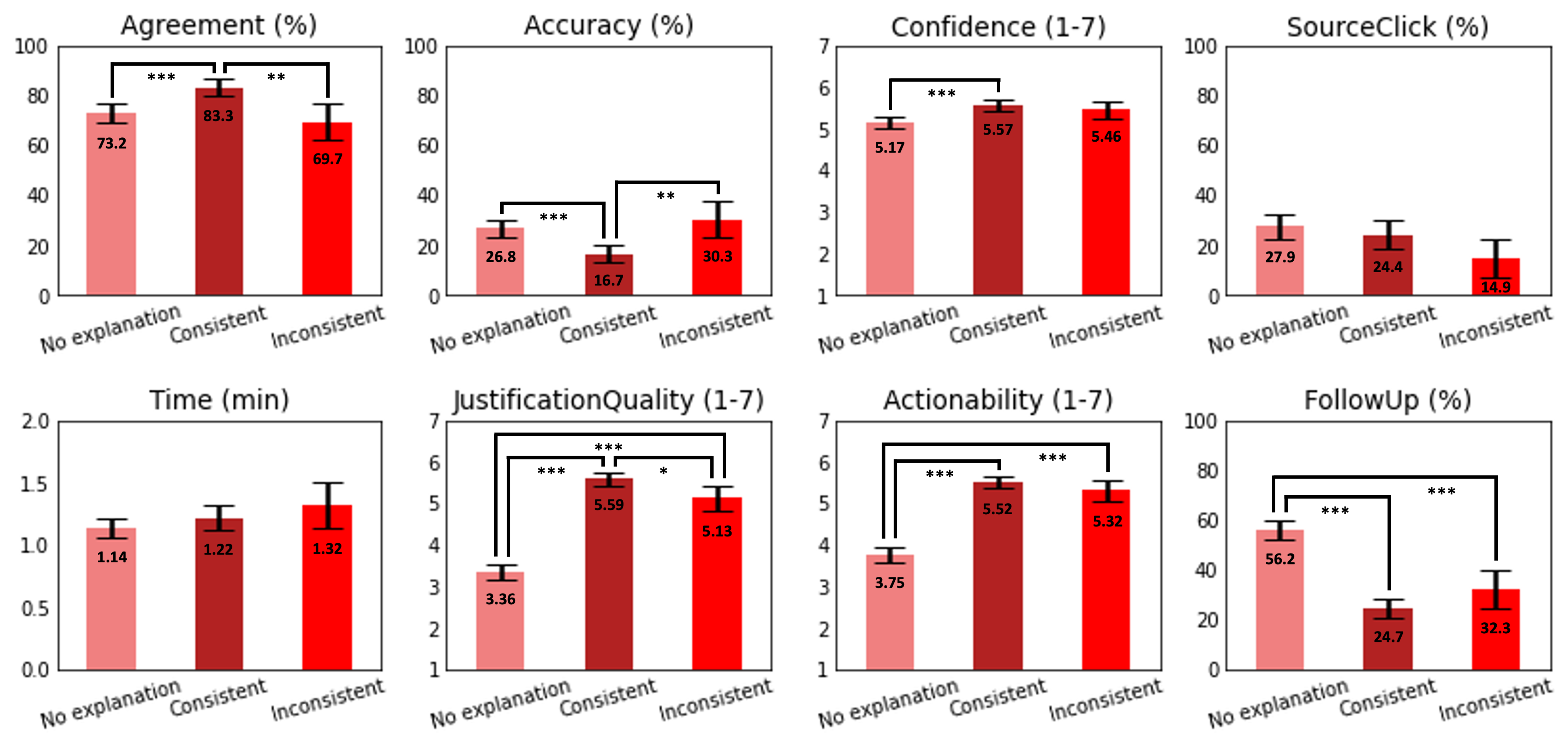}
\caption{\textbf{Study 2 results on inconsistencies.} We plot the raw data means and 95\% confidence intervals. Brackets indicate statistically significant differences between three types of incorrect LLM responses: No explanation, Consistent explanation, and Inconsistent explanation. Significance is marked as $^\ast$ ($p < .05$), $^{\ast\ast}$ ($p < .01$), and $^{\ast\ast\ast}$ ($p < .001$). See \cref{sec:inconsistencies} for details.}
\label{fig:inconsistencies}
\end{figure*}

\subsubsection{Additional effects of LLM accuracy}
\label{sec:mainaccuracy}

Finally, we find a significant main effect of LLM accuracy on many DVs, in addition to its interactions with explanation and sources reported above.
Provided that there are no sources or explanation, when the LLM answer is incorrect compared to correct, agreement is higher ($M = 78.2\%$ vs. $67.2\%$, $\beta = .60, SE = .19, p = .002$), confidence is higher ($M = 4.92$ vs. $4.55$, $\beta = .37, SE = .10, p < .001$), and \var{Actionability} is higher ($M = 2.91$ vs. $2.56$, $\beta = .35, SE = .13, p = .007$), while accuracy is lower ($M = 21.8\%$ vs. $67.2\%$, $\beta = -2.07, SE = .19, p < .001$) and time on task is lower ($M = .89$ min vs. $1.05$ min, $\beta = -.17, SE = .08, p = .025$).
These results suggest that participants found incorrect answers more plausible than correct answers for the task questions used in the experiment.
This is not surprising as we deliberately selected challenging questions, i.e., questions with less than 50\% human accuracy in our pilot study.
As such, this is likely a feature of our stimulus materials rather than a generalizable finding.

\subsection{Study 2 Results: Additional Analyses}
\label{sec:study2additional}

Finally, we report results from our additional pre-registered analyses on the effects of inconsistencies in explanations (\cref{sec:inconsistencies}) and the relationship between participants' source clicking behavior and other DVs (\cref{sec:sourceclick}).

\subsubsection{Inconsistencies in explanations}
\label{sec:inconsistencies}

In Study 1, we found inconsistencies in explanations to be an important unreliability cue that participants often noticed.
While the presence of inconsistencies is not something we control for or manipulate, we explore whether and how the natural inconsistencies that arose in LLM responses have effects on the DVs with a pre-registered analysis. 
Specifically, we used analysis of variance (ANOVA) to compare the means of DVs across three types of incorrect LLM responses: No explanation ($N = 616$), consistent explanation ($N = 461$), and inconsistent explanation ($N = 155$), where
$N$ indicates the number of instances for which participants received a given response type.
If there was a significant difference across response types, we conducted pairwise comparisons with post-hoc Tukey tests.
We only analyze responses with an incorrect answer, as none of the responses with a correct answer contained inconsistencies (as described in \cref{sec:study2materials}).
We present the results in \cref{fig:inconsistencies}.

For most DVs (all except \var{SourceClick} and \var{Time}) we find a significant difference across response types.
Overall, overreliance on incorrect answers is most prevalent when participants receive consistent explanations, as evidenced by the highest agreement with the LLM answer, confidence in their final answer, and ratings of justification quality and actionability, as well as the lowest accuracy and likelihood of asking follow-up questions.
In comparison, when participants receive inconsistent explanations, agreement is significantly lower ($M = 69.7\%$ vs. $83.3\%$ $p = .002$), rating of the LLM response's justification quality is significantly lower ($M = 5.13$ vs. $5.59$, $p = .028$), while accuracy is significantly higher ($M = 30.3\%$ vs. $16.7\%$, $p = .002$).
While our study materials did not allow us to investigate the effect of inconsistencies when an LLM answer is correct, which may happen less naturally based on our observations, these results suggest that \textbf{inconsistencies can help reduce overreliance on incorrect answers induced by explanations.}

Consistent with the quantitative results, 19 participants stated in their free-form answers that they disagreed with Theta 
when \shortquote{Theta's responses were contradictory.}
For example, several participants wrote about how Theta provided a logically inconsistent response to the question ``Do more than two thirds of South America's population live in Brazil?'' (See \cref{fig:task} for the response.)
As one participant elaborated, \shortquote{The Brazilian and South American population answer contradicted itself. Two-thirds would imply 66\% but given the number of Brazilians compared to the total population of South America given in the answer, the actual percentage is closer to 50\%.} \looseness=-1

\subsubsection{Source clicking behavior}
\label{sec:sourceclick}

From our main analysis (\cref{sec:study2results}), we did not find any significant effect on when participants chose to click on the provided source links.
We only found a marginally significant main effect of explanation such that participants' source clicking likelihood is lower when there is an explanation than not ($M = 25.0\%$ vs. $28.2\%$, $\beta = -.62, SE = .36, p = .086$).
However, there is high variance across individuals.
According to our tracking, 189 out of 308 participants never clicked on sources, 33 participants clicked on sources in one task, 18 participants in two tasks, 23 participants in three tasks, and 45 participants in all four tasks for which sources were provided.

To better understand participants' source clicking behavior, we conducted a pre-registered analysis to examine its relationship with other DVs.
Specifically, we used ANOVA to compare the means of DVs between instances in which participants were provided sources but did not click on any ($N = 914$) and instances in which participants were provided sources and clicked on one or more ($N = 318$).
Among the latter, 164 are instances in which the LLM answer is correct and 154 are instances in which the LLM answer is incorrect. \looseness=-1

We find that when participants click on sources, accuracy is higher ($M = 60.1\%$ vs. $49.2\%$, $p < .001$) and time on task is higher ($M = 2.11$ min vs. $1.08$ min, $p < .001$), while rating of the LLM response's justification quality is lower ($4.58$ vs. $5.08$, $p < .001$).
The accuracy and time on task results are intuitive.
The sources in our study stimuli tended to provide accurate and relevant information (see \cref{sec:study2materials} for details), so when participants clicked on sources, they likely found correct answers at the expense of spending more time on task.
Indeed, we see that source clicking was helpful when the LLM gave an incorrect answer.
The increase in accuracy is bigger when the LLM answer is incorrect ($M = 37.0\%$ vs. $24.2\%$) than when the LLM answer is correct ($M = 81.7\%$ vs. $74.8\%$).
For reference, when LLM responses do not include sources, participants' answer accuracy is $M = 19.5\%$ when the LLM answer is incorrect and $M = 72.7\%$ when the LLM answer is correct.

There are multiple possible factors that might influence the finding that the rating of justification quality is lower when participants have clicked on the provided sources. Participants may have clicked on sources because they found the LLM response's justification quality to be low, or their rating may have decreased after examining the sources.
Again breaking down the data into instances in which the LLM answer is correct and those where it is incorrect, participants' rating of the response's justification quality when they clicked on sources vs. not is $M = 5.04$ vs. $5.29$ when the LLM answer is correct and $M = 4.08$ vs. $4.87$ when the LLM answer is incorrect.

Together, these results suggest engaging with the content of (accurate and relevant) sources can be an effective way of improving decision outcomes.
However, the presence of explanation may reduce users' natural tendency to examine sources, especially when they find the explanation to be of high quality.
It could be helpful to nudge users to pay more attention to sources by highlighting sources or placing sources above explanations.

\section{Discussion}
\label{sec:discussion}

\subsection{Implications of Findings}

\subsubsection*{Explanations}
In our studies, we found that explanations play an important role in shaping users' reliance.
In Study 1, we gained qualitative insights on how participants interpreted and used explanations to judge the reliability of LLM answers.
In Study 2, we examined the effects of the presence of an explanation, as well as its interaction with other variables, and found that explanations increase reliance on both correct and incorrect responses.
This is consistent with prior findings in HCI that explanations can increase overreliance \cite{Bansal2021CHI,zhang2020effect,Poursabzi-Sangdeh-CHI2021,wang2021explanations,Fok2024Verifiability}, including explanations generated by LLMs~\cite{si2024fact,Pafla2024CHI}.
It is also consistent with prior work in psychology, which finds that explanations are often found compelling even when they contain little content \cite{langer1978mindlessness,Giffin2017} or content that experts judge irrelevant \cite{hopkins2016seductive}, and that effects of superficial cues on explanation quality are more severe when time and prior knowledge are limited \cite{hopkins2019,kelemen2013professional}. 
In the absence of effort and expertise, users will inevitably rely on superficial cues to explanation quality, such as fluency \cite{trout2008}, a characteristic that LLM explanations typically possess in spades.
This suggests a potential tension in providing LLM explanations to lay users: the properties that make such explanations intelligible and compelling may be precisely those that lead to overreliance.
As such, we encourage LLM explanations to be evaluated and optimized for appropriate reliance, in addition to other qualities such as fluency, justification quality, and satisfaction.

\subsubsection*{Sources}
Our results offer some basis for optimism, however: sources helped reduce overreliance on incorrect answers and increase appropriate reliance on correct answers.
One possibility is that sources encouraged participants to engage in slow and careful System 2 thinking, instead of quick and automatic System 1 thinking \cite{Kahneman2003, kahneman2011thinking}.
In our study, participants spent significantly more time on task when provided with sources, especially when the LLM's answer was incorrect.
The qualitative data also supports this. Many participants wrote that they checked sources. Many also wrote that they submitted a different answer from the LLM's answer when it conflicted with the information in the sources.
We emphasize, however, that the sources provided in Study 2 were all real and tended to provide accurate and relevant information.
This is not always the case. Recent work has found that popular LLM-infused applications frequently generate statements that are not supported by sources \cite{liu2023evaluating} and sometimes even generate fake sources \cite{Alkaissi2023}.
If the provided sources are junk or just broken links, then presumably they will not help foster appropriate reliance. They could potentially even hurt by making the LLM response look more trustworthy, similar to how flawed and meaningless explanations have been found to increase people's trust and reliance \cite{Eiband2019Placebic,Schemmer2022AIES,kaur2020CHI}.
In addition to improving the quality of sources in LLM responses, future work should explore different issues with sources (e.g., fake, unreliable, conflicting sources and inaccurate summaries of sources), design choices (e.g., location of sources and amount of preview), and their effects on people's perceptions and behaviors.

When it comes to choosing between providing sources only and providing sources and an explanation, there are benefits and drawbacks to each.
When the LLM answer is incorrect, participants' accuracy is highest on responses with sources only ($M = 31.8\%$), followed by responses with explanation and sources ($M = 23.1\%$), neither ($M = 21.8\%$), and explanation only ($M = 17.2\%$) --- suggesting that providing sources only is most effective at reducing overreliance on incorrect answers.
However, it is not as effective at improving appropriate reliance when the LLM answer is correct. Here, participants' accuracy is highest on responses with explanation and sources ($M = 79.9\%$), followed by responses with explanation only ($M = 78.2\%$), sources only ($M = 73.4\%$), and neither ($M = 67.2\%$).
In contexts where LLMs have much higher accuracy than users, providing sources only can lead to lower overall accuracy than providing sources and an explanation.
Further, participants rated responses with sources only lower in terms of justification quality and actionability, compared to responses with sources and explanation, suggesting that people prefer the latter.

\subsubsection*{Inconsistencies and other unreliability cues}
Finally, we found that LLM responses contain new forms of unreliability cues.
Prior research, in particular the work by \citet{chen2023understanding}, found that people identify AI models' biases, inability to consider contexts or multiple features, and lower performance on rare instances as cues of unreliability.
In our studies, we identified other cues such as inconsistencies in explanations and lack of explanation or sources --- all of which are related to the particular characteristics of LLMs.
For example, some inconsistencies occur due to the stochastic nature of LLMs: LLMs can generate different responses for the same input, unlike deterministic AI models.
Even within a single response, inconsistencies occur because LLMs are not trained to generate only logically consistent statements.
The other cues are connected to LLMs' natural language modality and ability to handle a wide variety of tasks, which lead to responses with much more diverse features and forms compared to classical AI models with fixed output spaces. \looseness=-1

Intriguingly, we found positive effects of such unreliability cues when it comes to reducing overreliance.
In Study 1, participants who noticed unreliability cues engaged with the LLM responses more thoroughly.
In Study 2, participants relied less on incorrect LLM responses when they were provided with explanations containing inconsistencies than those without.
These findings, along with prior findings on other unreliability cues (e.g., inconsistencies between multiple responses \cite{si2024fact,Lee2024FAccT}), suggest that guiding people's attention to these cues can be an effective approach to reducing overreliance.
For example, we could apply computational methods to automatically detect inconsistencies (e.g., ~\cite{tacl2022,contradiction2008}) then use highlighting to draw people's attention to the detected inconsistencies. Other interventions (e.g., expressing uncertainty, displaying disclaimers, and encouraging source checking) could be applied jointly for cases in which inconsistency detection is difficult or where LLM responses are consistently inaccurate.
We suggest future research to explore more thoroughly what unreliability cues exist in LLM responses and how to design interventions that help people notice and reason about these cues.

\subsection{Explanation of the Answer vs. Explanation of How the LLM Arrived at the Answer}

Throughout the work, we have used the term \textit{explanation} to refer to supporting details in LLM responses that justify the LLM's answer to the input question.
This is different from how the term is often used within the explainable AI community in that we do not make any assumptions about the extent to which it faithfully describes the way that the model arrived at its answer. 
We emphasize that faithfulness is extremely difficult for users --- or even model developers --- to evaluate, especially without access to the model's internals.
Evaluating the faithfulness of model explanations is an active area of research \cite{weijie2024,Zhao2024,atanasova2023faithfulness,jacovi2020faithfulness,agarwal2024faithfulness}.

Nevertheless, many participants in Study 1 interpreted ChatGPT's responses as including somewhat faithful explanations of how the system arrived at its answer, especially when the responses had certain characteristics \interview{3,6,7,8,10,11,14,15,16}.
For some, the critical characteristic was the presence of sources \interview{6,16}. As P16 (high-knowledge, low-use) described, \shortquote{I would think of the citation itself as an explanation because it kind of implies `I'm giving you this information because it came from this source' and then me as a human can evaluate that source.}
For others, it was the step-by-step form of responses, which are common for math questions \interview{3,7,8,14}. After seeing them, P14 (low-knowledge, high-use) said, \shortquote{I think it's very clear how did it [ChatGPT] provides me the answer.} \looseness=-1

In contrast, three participants, all with high knowledge of LLMs, were strongly opposed to the idea that ChatGPT could provide explanations of how it arrived at its answers \interview{5,12,13}.
P5 (high-knowledge, low-use) stated that \shortquote{it's provably false that ChatGPT's responses provide a description of how it arrives at its answers,} emphasizing that ChatGPT's responses are \shortquote{definitely and empirically not explanations because there's no reflection in the model.}
Similarly, P12 (high-knowledge, high-use) said they don't think of ChatGPT as explaining anything to them, and that ChatGPT was just \shortquote{trained to provide answers that look like an explanation because that's what we would find most useful.}
P4 (high-knowledge, low-use) shared this view and emphasized that \shortquote{there's no way to interpret how the answer came from.}
They noted that the explanations ChatGPT offers describe \shortquote{how a normal person would reach the answer,} and are not explanations of how ChatGPT arrives at its answers. \looseness=-1

In sum, while there was considerable variability between individuals, we found that many participants, especially those without much knowledge of LLMs, viewed ChatGPT's responses as including somewhat faithful explanations for how the system arrived at its answer.
This raises a concern because first, again, there is no reliable way for users or anyone to evaluate their faithfulness without access to the system's internals, and second, recent work has found explanations from LLMs are often not faithful to their process \cite{turpin2023language,Zhao2024,wiegreffe2022reframing,Marasovi2021FewShotSW,lyu2023faithful} and can easily be manipulated, e.g., to rationalize incorrect information \cite{pan2023on,buchanan2021truth,Zellers2019Fakenews}.
Such assumptions can be strengthened by the increasing anthropomorphization of LLMs and lead to inappropriate reliance \cite{Weidinger2022Risk,Shanahan2024Anthro,Cohn2024CHI}.
We suggest future research to explore strategies for improving people's understanding of LLMs \cite{long2020literacy,Annapureddy2024literacy} and study how they are connected to reliance behaviors. \looseness=-1

\subsection{Limitations}
\label{sec:limitations}

There are several limitations of our work that are worth reflecting on.
First and foremost, our studies were conducted in the context of objective question-answering and may not generalize to other contexts of LLM use (e.g., writing, idea generation, and task automation).
We encourage the community to conduct more empirical studies on how user reliance is shaped in various contexts.

Each of our studies has a different set of strengths and limitations.
Study 1 was a think-aloud study that offered descriptive examples of how users interpret and act upon different LLM response features in a relatively natural setting.
However, prior work has pointed out that the set-up of a think-aloud study can cause people to behave differently than they would otherwise \cite{Hertzum2009ScrutinizingUE,Boren2000,fox2011procedures}.
For example, we saw a much higher rate of source clicking in Study 1 (M = 63.8\%) than in Study 2 (M = 25.8\%) which was an online experiment. 
We also emphasize that the LLM response features identified in Study 1 are not comprehensive. We suggest future work to explore what other features influence users' reliance and can help them succeed in tasks despite inaccuracies from LLMs.

In Study 2, we employed a different research method (a controlled experiment), prioritizing the generalizability of findings by controlling as many other variables as possible.
For example, in the experiment, participants saw exactly one response from Theta, created in advance using the state-of-the-art LLM-infused applications ChatGPT and Perplexity AI, instead of interacting with a real system in multiple rounds.
While participants referred to Theta as ``AI'' or ``LLM'' in the exit questionnaire (e.g., \shortquote{I just trusted the AI when I didn't know the answer already}), we did not measure participants' general perceptions of Theta or inquire about their experience. 
Hence, it is more accurate to view Study 2 as a study of people's perceptions and behaviors around specific LLM responses rather than a study of people's interactions with LLMs.
While showing one controlled LLM response is a commonly used method (e.g.,~\cite{Kim2024FAccT,si2024fact,Lee2024FAccT}), people's perceptions and behaviors may change over time, meriting further studies in more interactive settings. \looseness=-1

Additionally, we set Theta's accuracy to be 50\% which is significantly worse than the state-of-the-art. While this choice allowed us to compare the effects of LLM response features on relying on correct vs. incorrect answers in a balanced fashion, future work should explore whether there are interaction effects between these features and the LLM's accuracy.
There are also implications of our experimental task, which was answering difficult factual questions (that less than 50\% of pilot study participants knew the answer to). 
We chose this task to simulate realistic scenarios of people seeking answers to questions they don't know the answer to. However, it is possible that our findings may not generalize to tasks where people have sufficient prior knowledge and can more deeply engage with the content of the LLM responses.
Finally, there are many LLM response features that we did not study or control for (e.g., simplicity of explanations \cite{lombrozo2007simplicity}, quality of sources \cite{Rieh2007Credibility,Wathen2002Credibility}, and presence of jargon \cite{cruz2024effect}).
We encourage future work to explore different features and methods to understand user interactions with LLMs, an emerging research area whose importance will only grow with time. \looseness=-1

\section{Conclusion}

We conducted two empirical studies to understand how different features of LLM responses shape users' reliance.
We found that the presence of explanations increases reliance on both correct and incorrect responses. 
However, we observed less reliance on incorrect responses when sources are provided or when explanations exhibit inconsistencies.
Our findings highlight the importance of evaluating LLM response features with users before deployment.
Our findings also suggest that providing (accurate and relevant) sources and designing interventions that help users notice and reason about inconsistencies and other unreliability cues in explanations can be promising directions for fostering appropriate reliance on LLMs.

\begin{acks}
We foremost thank the participants for sharing their time and experiences. We also thank the members of the Princeton Visual AI Lab, the Princeton HCI Lab, and the Princeton Concepts \& Cognition Lab, as well as the anonymous reviewers for thoughtful feedback and discussion. We acknowledge support from the NSF Graduate Research Fellowship Program (SK) and the Princeton SEAS Howard B. Wentz, Jr. Junior Faculty Award (OR).
\end{acks}

\bibliographystyle{ACM-Reference-Format}
\bibliography{references}

\appendix

\section*{Appendix}

\renewcommand{\thesection}{\Alph{section}}

\renewcommand\thefigure{\thesection\arabic{figure}}
\renewcommand\thetable{\thesection\arabic{table}}

The appendix is structured in the following way.
\begin{itemize}
    \item \textbf{\cref{app:study2followup}.} Additional Analyses: Study 2 Follow-up Questions
    \item \textbf{\cref{app:study1protocol}.} Study 1 Think-Aloud Protocol
    \item \textbf{\cref{app:study1questions}.} Study 1 Task Questions Used
    \item \textbf{\cref{app:study2participants}.} Study 2 Participant Demographics and LLM Background
    \item \textbf{\cref{app:study2wording}.} Study 2 Full Wording of the Experiment
    \item \textbf{\cref{app:study2stimuli}.} Study 2 Task Questions and LLM Responses Used
\end{itemize}

\section{Additional Analyses: Study 2 Follow-up Questions}
\label{app:study2followup}

As described in Section 4.1.2 of the main paper, in Study 2, we asked the participant if they would like to ask a follow-up to Theta.
We conducted two additional analyses to understand when and what types of follow-up questions participants asked.
First, we explored the the relationship between \var{FollowUp} (whether the participant writes a follow-up question they would like to ask) and \var{Agreement} (whether the participant submits the same answer as Theta's answer).
We found that the likelihood of writing a follow-up question when participants submit the same answer as Theta's answer vs. not is 33.4\% vs. 56.6\%. Breaking down the data into instances in which Theta's answer is correct and those where it is incorrect, the likelihood is 33.4\% vs. 46.5\% when Theta's answer is correct and 33.4\% vs. 67.5\% when Theta's answer is incorrect. That is, participants were more likely to ask follow-up questions when they disagreed with Theta's answer. \looseness=-1

We next analyzed what types of follow-up questions participants wrote.
In 43.8\% of these instances, LLM responses had neither an explanation nor sources; in 17.9\%, responses had an explanation only; in 26.1\%, responses had sources only; and in 12.3\% responses had both an explanation and sources.
To analyze the follow-up questions, we first went through 100 randomly sampled questions to develop codes, then another 100 to verify that the codes have saturated.
In total, we developed 11 codes which we grouped into 3 categories: (1) questions about related content, (2) requests for sources or explanations, and (3) expressions of doubt or disagreement.
We present them below along with their percentage out of 200 analyzed questions.
The percentages do not add up to 100 because some questions were coded into more than one category. \looseness=-1

\subsection*{(1) Questions about related content (61.0\%)}

The first category includes follow-up questions related to the task question.
Most frequent were \textbf{questions directly related to the LLM's answer (36.0\%)}. 
For example, while solving the task question ``Which animal was sent to space first, cockroach or moon jellyfish?'' one participant wrote \shortquote{What was the name and year of the NASA mission where a moon jellyfish was sent to space?} after receiving a response that only includes the date of when the first cockroaches were sent to space.
Also frequent were \textbf{questions that go beyond what is required to answer the task question (17.5\%)}. For example, while solving the task question ``Is it possible to scuba dive at the sunken city of Port Royal?'' one participant wrote \shortquote{When is good time of year to scuba dive at sunken city of Port Royal?}
Other questions were simpler.
Sometimes participants asked \textbf{why questions about the LLM's answer (4.5\%)}, e.g., one participant wrote \shortquote{Why is the brain larger now?}after receiving a response to the task question ``Is the human brain smaller or bigger than it was 100,000 years ago?''
Finally, while rare, some participants just re-asked the \textbf{task question (1.0\%)}, e.g., \shortquote{How long does food typically take to digest?} which is a short and open-ended version of the task question ``How long does it take for food to travel from our mouth to our stomach in general, 7 seconds or 30 seconds?''

\subsection*{(2) Requests for sources or explanations (28.5\%)}

The second category concerns requests.
To begin, many participants used the follow-up question box to request \textbf{sources (17.5\%)}, or links, evidence, or proof, e.g., \shortquote{What sources are you using for your answer?}
Going beyond this, some participants requested \textbf{quotes or summaries of sources (2.0\%)}, e.g., \shortquote{Using your sources, find quotes that support the correct answer.}
Participants also made requests for an \textbf{explanation of the answer (5.5\%)}, or additional details, information, justification, facts, or statistics, e.g., \shortquote{Can you give me some reasons as to why this could be true?}
Finally, participants requested for an \textbf{explanation of how the LLM arrived at the answer (3.5\%)}, e.g., \shortquote{How did you make this determination?} where the emphasis is placed on how the LLM knows the answer, not necessarily why the answer is correct. \looseness=-1

\subsection*{(3) Expressions of doubt or disagreement (14.5\%)}

The final category concerns questions expressing doubt or disagreement.
These include questions that point out \textbf{inconsistencies in the LLM's explanations (1.5\%)}, e.g., \shortquote{213 isn't two thirds of 430. Wouldn't that be less than two thirds?} as well as \textbf{gaps and contradictions between the LLM's answer and the sources (3.5\%)}, e.g., \shortquote{None of those sources talk about gorillas, where did you get your info from?} and \shortquote{Why did one of your sources disagree with your statement?}
In other examples, participants challenged the LLM's answer for unspecified reasons, likely based on their \textbf{intuition (9.5\%)}, e.g., \shortquote{Ok, that can't be right. Are you sure?} and \shortquote{A million sounds astronomical to be honest. There is no way you are correct here.} \looseness=-1

\subsection*{Discussion}

Together these findings suggest that participants desired to see sources and explanations in LLM responses and tended to ask follow-up questions when they had doubts or disagreed with the LLM's answers. These are consistent with the quantitative findings presented in the main paper: the presence of sources and explanations significantly decreased the likelihood of asking a follow-up question. We also observed that ratings of the LLM response's justification quality and actionability tend to be negatively correlated with the likelihood of asking a follow-up question. 
In our experiment, we did not show participants the LLM's responses to their follow-up questions so that we can fully control the number and content of responses.
Exploring the effects of follow-up interactions would be an important future research direction. \looseness=-1

\section{Study 1 Think-Aloud Protocol}
\label{app:study1protocol}

We conducted the think-aloud study based on the following questions.

\subsubsection*{Background}
\begin{itemize}
    \item How would you describe your knowledge of artificial intelligence or AI?
    \item How would you describe your knowledge of LLMs and LLM-infused applications such as ChatGPT, Copilot, and Gemini? 
    \item Do you use any LLMs and LLM-infused applications? If so, what do you use? How frequently do you use it and for what tasks?  What is your main reason for using it? How satisfied are you with it overall?
\end{itemize}

\subsubsection*{Setup \& Introduction}
\begin{itemize}
    \item Now I will ask you to complete a few tasks with ChatGPT (an LLM-based chatbot developed by OpenAI). Please open your browser and go to \url{https://chat.openai.com/}. We have created an account for this study. Please use this login information to sign into the account. 
    \item {[}Show three tasks.{]} These are the tasks I want you to complete using ChatGPT. They are not trick questions, and all have a correct answer. Please let me know if you already know the answer to any of the questions off the top of your head.
    \item For each question, I would like you to first, try to find the correct answer using ChatGPT, and second, indicate how confident you are in the correctness of the answer you ended up with. 
\end{itemize}

\subsubsection*{Part 1 (Base)}
\begin{itemize}
    \item To get started, ask the first task question to ChatGPT. Please read its response and write follow-up messages as needed. Feel free to start as many new chats as you want, and please start a new chat for a new task question. Please think aloud while completing the task. That is, try to say everything that comes to your mind while you engage with the task.
    \item {[}For each task, ask the following questions if the participant does not think aloud.{]}
    \begin{itemize}
        \item What do you think is the correct answer to the task question?
        \item How confident are you in your answer on a scale of 1 to 7?
        \item What would help you gain more confidence in your answer?
    \end{itemize}
    \item {[}After all three tasks, ask the following questions.{]}
    \begin{itemize}
        \item Do you view ChatGPT's responses as explanations? What counts as explanations to you? 
        \item How do you think ChatGPT generates responses?
        \item Do you think ChatGPT's responses provide a description of how it arrived at its answer to the question?
        \item Would you find it useful to know how ChatGPT arrives at its answer?
    \end{itemize}
\end{itemize}

\subsubsection*{Part 2 (Prompting)}
\begin{itemize}
    \item Now I am going to ask you to complete the tasks again. But this time, try following-up on the provided responses Here are some example prompts you can use, but be creative! Feel free to try as many.
    \item {[}Show prompt examples. The order is randomized for each participant.{]}
    \begin{itemize}
        \item I think you are wrong. Try again.
        \item Are you sure about [a specific part of ChatGPT’s response]?
        \item Explain if [a specific part of the answer] is correct.
        \item Explain why your answer may be wrong.
        \item Explain why the correct answer is [opposite from ChatGPT’s previous answer]? 
        \item Can you explain in a different way?
        \item Explain step by step.
        \item Provide a shorter explanation.
        \item Explain like I’m five.
        \item Are there other reasons for the answer?
        \item Provide an explanation with supporting sources.
        \item Explain how confident you are in the answer.
    \end{itemize}
    \item {[}For each task, ask the following questions if the participant does not think aloud.{]}
    \begin{itemize}
        \item What do you think is the correct answer to the task question?
        \item How confident are you in your answer on a scale of 1 to 7?
        \item What would help you gain more confidence in your answer?
    \end{itemize}
    \item {[}After all three tasks, ask the following questions.{]}
    \begin{itemize}
        \item Have there been any changes in your thoughts about ChatGPT?
        \item What did you think about the prompts?
        \item Is there anything that you want the research team to know that we haven’t been able to cover yet?
    \end{itemize}
\end{itemize}

\section{Study 1 Task Questions Used}
\label{app:study1questions}

As discussed in Section 3.1.1 of the main paper, each participant was given three questions: a general domain factual question, a health or legal domain factual question, and a math question. The specific question was randomly selected from the following.

\begin{itemize}
    \item General domain factual questions
    \begin{itemize}
        \item Has Paris hosted the Summer Olympics more times than Tokyo?
        \item Did Tesla debut its first car model before or after Dropbox was founded?
        \item Is it cheaper to buy three stocks of Moderna than two stocks of Pepsi?
    \end{itemize}
    \item Health or legal domain factual questions
    \begin{itemize}
        \item Can you get hepatitis A after having hepatitis B?
        \item Can one get Lyme disease from mosquitoes?
        \item Can a dead wasp inject one with venom?
        \item Is it illegal to collect rainwater in Colorado?
        \item Can one legally own a giraffe in Oregon without a permit?
        \item Can one be charged with a crime for stealing wifi in California?
     \end{itemize}
    \item Math questions
    \begin{itemize}
        \item Jessica is six years older than Claire. In two years, Claire will be the same age as Millie Bobby Brown now. The age of Claire's father twenty years ago is 3 times the current age of Jessica. How old is Claire’s father now?
        \item Sue puts one grain of rice on the first square of a Go board and puts double the amount on the next square. How many grains of rice does Sue put on the last square?
        \item A bird watcher records the number of birds he sees each day. The number of birds he saw on Monday is equivalent to the 12th Fibonacci number. On Tuesday he sees half as many birds as he did on Monday. On Wednesday he sees 2 more birds than he did on Tuesday. How many total birds did the bird watcher see from Monday to Wednesday?
    \end{itemize}
\end{itemize}

\section{Study 2 Participant Demographics and LLM Background}
\label{app:study2participants}

In the exit questionnaire, participants optionally self-reported their gender, age, race, ethnicity, and level of education.
Of 308 participants in the final sample, 38.3\% identified as woman, 58.8\% as man, 1.9\% as non-binary, 0.3\% as gender-diverse, and 0.3\% as transgender.
The age distribution was as follows: 18--24 (9.1\%), 25--34 (38.0\%), 35--44 (26.0\%), 45--54 (15.9\%), 55-64 (7.1\%), and 65--74 (2.6\%).
Regarding race, 56.5\% identified as white alone, followed by Black or African American alone (17.2\%), Asian alone (7.8\%), two or more races (7.1\%), American Indian or Alaska Native alone (1.3\%), and Middle Eastern or North African alone (0.6\%).
Regarding ethnicity, 8.8\% identified as Hispanic or Latino.
For the highest level of education completed, 38.0\% said 4 year degree, followed by some college (19.2\%), professional degree (14.3\%) and high school graduate (14.3\%), 2 year degree (9.4\%), doctorate (2.6\%), less than high school (1.6\%).

We also measured the following variables based on participants' self-reported ratings on a 5-point scale:
\begin{itemize}
    \item \var{LLM\_Knowledge}: Rating on the question ``How familiar are you with LLMs and LLM-infused applications such as ChatGPT, Copilot, and Gemini?'' 
    \item \var{LLM\_Use}: Rating on the question ``How often do you use LLMs and LLM-infused applications such as ChatGPT, Copilot, and Gemini?''
    \item \var{LLM\_Attitude}: Rating on the question ``Overall, how do you feel about LLMs and LLM-infused applications such as ChatGPT, Copilot, and Gemini?'' 
\end{itemize}

The mean and standard deviation of \var{LLM\_Knowledge} are $3.02 \pm 0.88$ around ``3: Moderately familiar, I know what they are and can explain.''
For \var{LLM\_Use}, they are $3.04 \pm 1.30$ around ``3: Sometimes, about 3--4 times a month.''
For \var{LLM\_Attitude}, they are $3.70 \pm 1.03 $ between ``3: Neutral'' and ``4: Somewhat positive.''

\section{Study 2 Full Wording of the Experiment}
\label{app:study2wording}

In this section, we present the full wording of the experiment. See Section 4.1.1 for a description of the study procedure.

\subsubsection*{About the study}
\begin{itemize}
    \item Imagine you have a question that you don't know the answer to. You have access to an AI system that is connected to the internet and can answer a wide range of questions, so you've asked your question and received a response. What will you do now?

    \item In this study, you will answer a set of questions with the help of an AI system prototype named ``Theta.'' You will read more about Theta in the next page. In total, this study will take around 15 minutes to complete. Please complete everything in one sitting.
\end{itemize}

\subsubsection*{About the AI system}
\begin{itemize}
    \item The AI system prototype used in this study, ``Theta,''  is based on a ``large language model'' (LLM). When asked questions or given instructions, LLMs can generate human-like responses. These generated responses can sound convincing and fluent, but may not always be correct. LLMs can be used for many tasks, including question answering, text summarization, creative writing, and programming. You may have used or heard about some popular applications using LLMs such as OpenAI’s ChatGPT, Microsoft’s Copilot, and Google’s Gemini. Some are connected to the internet, while others are not.
    \item Theta uses similar technology to OpenAI’s ChatGPT. It is connected to the internet, and can answer a wide range of questions. However, as with all AI systems, its responses may be inaccurate, incomplete, and inconsistent, even when they sound convincing.
\end{itemize}

\subsubsection*{Are you ready to begin?}
\begin{itemize}
    \item You will now be asked to answer 8 questions, and you will have responses from Theta to help you decide how to answer. You may click on source links in Theta’s responses, but please do not conduct your own internet search. When you are ready to proceed, click on the “next” button.
\end{itemize}

\subsubsection*{Task (repeated 8 times)}
See Figure 2 in the main paper for screenshots of the task.

\subsubsection*{Post-task attention check}
\begin{itemize}
    \item Thank you for completing all the tasks! We're curious how well you remember the questions you were asked. For each question, indicate whether you were asked or not asked the question. [Display 10 randomly selected questions (around half asked and half not asked in the experiment).]
\end{itemize}

\subsubsection*{Open-ended questions}
\begin{itemize}
    \item If you submitted a final answer different from Theta’s answer, can you explain the reason? Please write 1–3 sentences.
    \item Please explain in 1–3 sentences how you rated the statement ``Theta's response offers good justification for its answer.''
    \item Please explain in 1–3 sentences how you rated the statement: ``Theta’s response includes information that helps me determine what my final answer should be.''
\end{itemize}

\subsubsection*{LLM background}
\begin{itemize}
    \item Note that these questions are about large language models (LLMs) and LLM-infused applications in general, not about the specific AI system ``Theta'' used in this study.
    \item How familiar are you with LLMs and LLM-infused applications such as ChatGPT, Copilot, and Gemini?
    \begin{itemize}
        \item Options: Not familiar at all, I have never heard of them; Slightly familiar, I have heard of them or have some idea of what they are; Moderately familiar, I know what they are and can explain; Very familiar, I have technical knowledge of what they are and how they work; Extremely familiar, I consider myself an expert on them
    \end{itemize}
    \item How often do you use LLMs and LLM-infused applications such as ChatGPT, Copilot, and Gemini?
    \begin{itemize}
        \item Options: Never; Rarely, about 1--2 times a month; Sometimes, about 3--4 times a month; Often, about twice a week; Always, about once or more a day
    \end{itemize}
    \item Overall, how do you feel about LLMs and LLM-infused applications such as ChatGPT, Copilot, and Gemini?
    \begin{itemize}
        \item Options: Negative; Somewhat negative; Neutral; Somewhat positive; Positive
    \end{itemize}
\end{itemize}

\subsubsection*{Demographics}
\begin{itemize}
    \item What is your age? 
    \item What is the highest degree of education you have completed? (If you’re currently enrolled in school, please indicate the highest degree you have received.)
    \begin{itemize}
        \item Options: Less than high school; High school graduate; Some college; 2 year degree; 4 year degree; Professional degree; Doctorate; Prefer not to answer
    \end{itemize}
    \item What gender do you identify with? (Select all that apply)
    \begin{itemize}
        \item Options: Woman; Man; Non-binary; Gender-diverse; Transgender; Prefer to self-describe; Prefer not to answer
    \end{itemize}
    \item How would you describe your race/ethnicity? (Select all that apply) 
   \begin{itemize}
       \item Options: American Indian or Alaska Native; Asian; Black or African American; Hispanic or Latino; Middle Eastern or North African; Native Hawaiian or Other Pacific Islander; White; Prefer to self-describe; Prefer not to answer
   \end{itemize}
\end{itemize}

\subsubsection*{Debrief}
\begin{itemize}
    \item Thank you for your participation in the study! To create a realistic experimental setting, we showed AI answers that are directly from, or only slightly adapted from, responses from an actual AI system. As known, AI systems can make up information. Please note that the AI answers you saw in this study may have been inaccurate, incomplete, or inconsistent, even when they sounded convincing. 
\end{itemize}

\section{Study 2 Task Questions and LLM Responses Used}
\label{app:study2stimuli}

As discussed in the main paper, we varied three variables in LLM responses: accuracy of the LLM's answer to the question, presence of an explanation, and presence of clickable sources. In total, there are 8 types of responses. In this section, for each of the 12 task questions, we show 2 responses, one with a correct answer and one with an incorrect answer, with both an \textcolor{brown}{explanation} and \textcolor{teal}{sources}. See Section 4.1.4 and Figure 3 in the main paper for how the other 6 types of responses would look like.

\subsubsection*{Which animal was sent to space first, cockroach or moon jellyfish?}
\begin{itemize}[leftmargin=*]
    \item \textbf{Correct}: A moon jellyfish was sent to space before a cockroach. \textcolor{brown}{The moon jellyfish was among the first animals sent to space as part of experiments to understand how microgravity affects biological organisms. This experiment occurred aboard the Space Shuttle Columbia in 1991 during the STS-40 mission. Scientists selected moon jellyfish because their simple structure and the way they navigate using gravity-sensitive cells made them ideal for studying the effects of zero gravity on orientation and movement. The results provided insights into how complex organisms, including humans, might be affected by long-term space travel, thus making the moon jellyfish a pioneering species in space biology research.} \textcolor{teal}{Sources: 1. \url{https://www.reddit.com/r/todayilearned/comments/j4wyzx/til_that_in_early_1990s_nasa_sent_jellyfish_to/} 2. \url{https://www.rmg.co.uk/stories/topics/what-was-first-animal-space} 3. \url{https://www.wired.com/2007/10/cockroach-birth/}}
    \item \textbf{Incorrect}: A cockroach was sent to space before a moon jellyfish, \textcolor{brown}{marking its status as one of the early non-human animals to experience space travel. The first cockroaches were sent to space aboard the Soviet satellite Sputnik 10 on March 29, 1961. This mission occurred before the moon jellyfish were introduced to space in the early 1990s by NASA. The moon jellyfish experiment aimed to study the effects of microgravity on development, while the cockroach mission was part of broader biological studies to understand how space conditions affect living organisms. Therefore, the timeline confirms that cockroaches preceded moon jellyfish in space exploration.} \textcolor{teal}{Sources: 1. \url{https://www.diversdirect.com/w/howling-at-the-moon-jellies} 2. \url{https://www.space.com/apollo-11-moon-dust-cockroaches-auction-withdrawn} 3. \url{https://en.wikipedia.org/wiki/Nadezhda_\%28cockroach\%29}}
\end{itemize}

\subsubsection*{Have more people been to the surface of the moon or the bottom of the Mariana Trench?}
\begin{itemize}[leftmargin=*]
    \item \textbf{Correct}: More people have visited the bottom of the Mariana Trench than have walked on the moon. \textcolor{brown}{As of recent counts, over 20 individuals have reached the trench's deepest point, known as Challenger Deep, due to advancements in deep-sea submersible technology and increased interest in ocean exploration. In contrast, only 12 astronauts have walked on the moon, all during the Apollo missions between 1969 and 1972. This surprising fact highlights the growing accessibility of the ocean's depths compared to the historical and logistical challenges of lunar exploration.} \textcolor{teal}{Sources: 1. \url{https://www.britannica.com/story/how-many-people-have-been-to-the-moon} 2. \url{https://en.wikipedia.org/wiki/List_of_people_who_descended_to_Challenger_Deep} 3. \url{https://www.defense.gov/News/Feature-Stories/story/Article/1737193/hitting-bottom-submariner-explored-deepest-part-of-ocean/}}
    \item \textbf{Incorrect}: More people have been to the surface of the Moon than to the bottom of the Mariana Trench. \textcolor{brown}{Twelve astronauts have walked on the lunar surface during the six Apollo missions between 1969 and 1972. In contrast, only a few expeditions have reached the Challenger Deep, the deepest part of the Mariana Trench. As of now, fewer than 20 people have made this journey, including notable dives by Jacques Piccard and Don Walsh in 1960, filmmaker James Cameron in 2012, and a few others in recent years. Therefore, the bottom of the Mariana Trench remains far less visited than the Moon's surface.} \textcolor{teal}{Sources: 1. \url{https://simple.wikipedia.org/wiki/List_of_people_who_have_walked_on_the_Moon} 2. \url{https://science.nasa.gov/moon/moon-walkers/} 3. \url{https://www.discovermagazine.com/planet-earth/what-has-been-found-in-the-deep-waters-of-the-mariana-trench}}
\end{itemize}

\subsubsection*{Which body part has a higher percentage of water, lungs or skin?}
\begin{itemize}[leftmargin=*]
    \item \textbf{Correct}: Lungs have a higher percentage of water compared to the skin \textcolor{brown}{because they are composed primarily of spongy, elastic tissue filled with air sacs (alveoli) that require moisture to function effectively. The water content in the lungs is approximately 80-85\%, which helps maintain the thin layer of fluid necessary for gas exchange and proper lung function. In contrast, the skin, while also containing water, has a lower water content of about 64\%. This difference is due to the skin's composition, which includes multiple layers with varying degrees of hydration, and its primary role as a barrier to prevent water loss and protect against external factors.} \textcolor{teal}{Sources: 1. \url{https://www.medicalnewstoday.com/articles/167533} 2. \url{https://pubmed.ncbi.nlm.nih.gov/7891399/} 3. \url{https://www.medicalnewstoday.com/articles/what-percentage-of-the-human-body-is-water}}
    \item \textbf{Incorrect}: Skin has a higher percentage of water compared to the lungs. \textcolor{brown}{The skin, being the largest organ of the body, is composed of approximately 64\% water. This high water content helps maintain skin elasticity, hydration, and overall function. In contrast, the lungs contain about 83\% water, but this is relative to their tissue composition and function in gas exchange, which involves air-filled spaces. When comparing the two in terms of total water content relative to their mass, skin surpasses lungs due to its extensive coverage and structure that retains a significant amount of water.} \textcolor{teal}{Sources: 1. \url{https://svalbardi.com/blogs/water/percentage} 2. \url{https://www.ncbi.nlm.nih.gov/pmc/articles/PMC6850423/} 3. \url{https://www.usgs.gov/special-topics/water-science-school/science/water-you-water-and-human-body}}
\end{itemize}

\subsubsection*{Do gorillas have twice as many hairs per square inch as humans?}
\begin{itemize}[leftmargin=*]
    \item \textbf{Correct}: No, gorillas don't have twice as many hairs per square inch as humans. \textcolor{brown}{Humans and gorillas have similar hair density per square inch. Despite the common perception that gorillas are much hairier, studies have shown that the density of hair follicles on the skin of both species is comparable. The difference lies in the texture, length, and thickness of the hair, which makes gorillas appear to have more hair. Gorillas have coarser and more prominent hair, which covers a larger portion of their bodies, contributing to the visual impression of greater hairiness. Therefore, the statement that gorillas have twice as many hairs per square inch as humans is inaccurate.} \textcolor{teal}{Sources: 1. \url{https://louisvillezoo.org/gorilla-forest/gorilla-facts/} 2. \url{https://www.nationalgeographic.com/science/article/the-semi-naked-ape-or-why-peach-fuzz-makes-it-harder-for-parasites} 3. \url{https://www.healthline.com/health/hair-density}}
    \item \textbf{Incorrect}: Yes, gorillas have twice as many hairs per square inch as humans. \textcolor{brown}{Gorillas have a significantly higher hair density compared to humans. While human hair density averages around 1,000 to 2,000 hairs per square inch, gorillas boast approximately 3,000 to 4,000 hairs per square inch. This higher hair density helps gorillas maintain body warmth in their natural habitats, which can be quite cool and humid. Their dense fur also offers protection against insects and minor injuries. Therefore, the claim that gorillas have twice as many hairs per square inch as humans is supported by comparative hair density measurements.} \textcolor{teal}{Sources: 1. \url{https://www.healthline.com/health/how-many-hairs-on-a-human-head} 2. \url{https://www.economist.com/science-and-technology/2011/12/17/the-not-so-naked-ape} 3. \url{https://bionumbers.hms.harvard.edu/bionumber.aspx?id=101509}}
\end{itemize}

\subsubsection*{Do all mammals except platypus give birth to live young?}
\begin{itemize}[leftmargin=*]
    \item \textbf{Correct}: No, there are mammals other than platypus that don't give birth to live young. \textcolor{brown}{The question overlooks that there are two main groups of mammals: monotremes and therians. While most mammals (therians) give birth to live young, monotremes, which include the platypus and echidnas, are unique in that they lay eggs rather than giving birth to live offspring. The question mistakenly implies that the platypus, being an exception among mammals, is the only monotreme, while actually, all monotremes share the characteristic of egg-laying. Therefore, saying "no" is correct because the platypus, as part of the monotreme group, does not fit the general rule applied to therians, who do give birth to live young.} \textcolor{teal}{Sources: 1. \url{https://www.earth.com/earthpedia-articles/are-there-mammals-that-lay-eggs/} 2. \url{https://animaldiversity.org/accounts/Monotremata/} 3. \url{https://www.britannica.com/animal/monotreme}}
    \item \textbf{Incorrect}: Yes, all mammals except platypus give birth to live young. \textcolor{brown}{With the exception of the platypus (which is one of the few monotremes), all other mammals are placental or marsupial and give birth to live young. Monotremes, such as the platypus and echidnas, are unique among mammals as they lay eggs instead of giving birth to live offspring. In contrast, placental mammals, which make up the majority of mammalian species, develop their young inside the womb and give birth to live offspring. Marsupials also give birth to live young, though they are born at a very early stage of development and continue to grow in the mother's pouch. Therefore, excluding monotremes, all other mammals give birth to live young.} \textcolor{teal}{Sources: 1. \url{https://australian.museum/learn/animals/mammals/platypus/} 2. \url{https://www.britannica.com/animal/mammal/Reproduction} 3. \url{https://www.ck12.org/flexi/life-science/mammal-reproduction/what-is-an-example-of-a-mammal-that-does-not-give-birth-to-live-young/}}
\end{itemize}

\subsubsection*{Do our eyes have more than a million moving parts?}
\begin{itemize}[leftmargin=*]
    \item \textbf{Correct}: Yes, our eyes have more than a million moving parts. \textcolor{brown}{The human eye's intricate structure includes numerous components that move to facilitate vision. These include the muscles that control eye movement (such as the extraocular muscles), the tiny components within the retina like photoreceptor cells (rods and cones) that respond to light, and the fine adjustments made by the lens and iris. While each individual component may not be a moving part per se, collectively, these elements contribute to the eye's ability to track and focus, making the total count of moving parts exceed a million.} \textcolor{teal}{Sources: 1. \url{https://2020visioncare.com/the-eye-a-marvel-of-complexity-with-over-2-million-working-parts/} 2. \url{https://midwesteyecenter.com/what-are-the-makings-of-the-human-eye/} 3. \url{https://www.optometrists.org/general-practice-optometry/guide-to-eye-health/how-does-the-eye-work/}}
    \item \textbf{Incorrect}: No, our eyes do not have more than a million moving parts. \textcolor{brown}{While the eye is a highly complex organ with several intricate components, including the cornea, lens, iris, and retina, the total number of distinct moving parts does not reach a million. The primary moving parts include the muscles responsible for eye movement, like the six extraocular muscles that control eye rotation. Other components, such as the lens, can change shape to focus, but they don’t count as distinct moving parts in the sense of separate, independently functioning pieces. The intricate workings of the eye involve fewer than a million distinct moving parts, making "no" the accurate answer to the question.} \textcolor{teal}{Sources: 1. \url{https://www.aao.org/eye-health/anatomy/parts-of-eye} 2. \url{https://2020visioncare.com/the-eye-a-marvel-of-complexity-with-over-2-million-working-parts/} 3. \url{https://en.wikipedia.org/wiki/Human_eye}}
\end{itemize}

\subsubsection*{Is the human brain smaller or bigger than it was 100,000 years ago?}
\begin{itemize}[leftmargin=*]
    \item \textbf{Correct}: The human brain is smaller now than it was 100,000 years ago. \textcolor{brown}{This reduction in size is thought to be related to evolutionary changes that occurred with the development of complex social structures and technologies. Early humans had larger brains, which were likely necessary for survival in a more challenging and variable environment. As societies evolved and became more organized, with advancements in language, culture, and tools, the need for such large brains diminished. Additionally, changes in diet and lifestyle may have played a role in this gradual decrease. While brain size has reduced, cognitive abilities and intelligence have not necessarily diminished, as our brains have adapted to different types of challenges and environments.} \textcolor{teal}{Sources: 1. \url{https://www.frontiersin.org/journals/ecology-and-evolution/articles/10.3389/fevo.2021.742639/full} 2. \url{https://geneticliteracyproject.org/2024/05/29/our-brains-are-getting-smaller-over-the-last-100000-years-whats-going-on/} 3. \url{https://www.bbc.com/future/article/20220503-why-human-brains-were-bigger-3000-years-ago}}
    \item \textbf{Incorrect}: The human brain is bigger today than it was 100,000 years ago. \textcolor{brown}{Over the course of human evolution, our brain size has increased significantly. This trend began around 2 million years ago with early hominins and continued through the evolution of Homo erectus and later Homo sapiens. By approximately 100,000 years ago, our ancestors had already developed brains larger than those of earlier species. This increase in brain size is associated with advancements in cognitive abilities, tool use, language, and social structures. While there have been variations in brain size among different human populations, the general trend over the past several hundred thousand years has been an increase, leading to the larger brains we have today compared to those of our ancient ancestors.} \textcolor{teal}{Sources: 1. \url{https://www.britannica.com/science/human-evolution/Increasing-brain-size} 2. \url{https://www.sciencedaily.com/releases/2022/08/220807102043.htm} 3. \url{https://humanorigins.si.edu/human-characteristics/brains}}
\end{itemize}

\subsubsection*{Do more than two thirds of South America's population live in Brazil?}
\begin{itemize}[leftmargin=*]
    \item \textbf{Correct}: No, not more than two-thirds of South America's population live in Brazil. \textcolor{brown}{Although Brazil is the largest country in South America by both land area and population, it does not encompass the majority of the continent's people. As of recent estimates, Brazil's population is roughly 213 million, while South America's total population is about 440 million. This means Brazil's population constitutes roughly 48\% of the continent's total, which is significantly less than two-thirds. Hence, while Brazil has a large population, it does not exceed the two-thirds threshold relative to the entire continent's population.} \textcolor{teal}{Sources: 1. \url{https://www.cia.gov/the-world-factbook/countries/brazil/} 2. \url{https://www.macrotrends.net/global-metrics/countries/BRA/brazil/population} 3. \url{https://www.worldometers.info/world-population/south-america-population/}}
    \item \textbf{Incorrect}: Yes, more than two-thirds of South America's population live in Brazil \textcolor{brown}{because Brazil is the largest and most populous country on the continent. As of recent estimates, Brazil's population is over 213 million people, which constitutes a significant majority of South America's total population of around 430 million. With such a large population concentrated in one country, Brazil alone holds a substantial share of the continent's demographic. Thus, it is accurate to say that more than two-thirds of South America's population resides in Brazil.} \textcolor{teal}{Sources: 1. \url{https://www.worldometers.info/world-population/brazil-population/} 2. \url{https://worldpopulationreview.com/continents/south-america} 3. \url{https://en.wikipedia.org/wiki/List_of_South_American_countries_by_population}}
\end{itemize}

\subsubsection*{Are all people born with fingerprints?}
\begin{itemize}[leftmargin=*]
    \item \textbf{Correct}: No, not all people are born with fingerprints. \textcolor{brown}{A condition called adermatoglyphia results in the absence of fingerprints. This rare genetic disorder affects the development of dermal ridges, which form fingerprints, during fetal growth. Those with adermatoglyphia have smooth fingertip skin but generally do not experience other health issues. Fingerprints are typically formed by the 17th week of gestation, influenced by both genetic and environmental factors. Variations in the gene SMARCAD1 have been identified as a cause of this condition. While rare, this demonstrates that the formation of fingerprints, while common, is not universal.} \textcolor{teal}{Sources: 1. \url{https://www.bbc.com/news/world-asia-55301200} 2. \url{https://www.nbcnews.com/health/body-odd/why-some-people-dont-have-fingerprints-flna1c6437284} 3. \url{https://www.smithsonianmag.com/science-nature/adermatoglyphia-genetic-disorder-people-born-without-fingerprints-180949338/}}
    \item \textbf{Incorrect}: Yes, all humans are born with fingerprints \textcolor{brown}{due to the way they develop during fetal growth. Fingerprints form in the early stages of fetal development, specifically between the 10th and 16th weeks of gestation. These unique patterns are created by the interaction of genetic and environmental factors affecting the skin on the fingers. Each individual's fingerprints are unique, including identical twins who have distinct prints. Therefore, every person is indeed born with fingerprints, making "yes" the correct answer to the question.} \textcolor{teal}{Sources: 1. \url{https://medlineplus.gov/genetics/understanding/traits/fingerprints/} 2. \url{https://www.tensor.co.uk/blog/news/10-facts-about-fingerprints} 3. \url{https://sites.rutgers.edu/fingerprinting/no-two-finger-prints-are-alike/}}
\end{itemize}

\subsubsection*{What type of tear is produced in larger quantities, basal tears or reflex tears?}
\begin{itemize}[leftmargin=*]
    \item \textbf{Correct}: Reflex tears are produced in larger quantities compared to basal tears \textcolor{brown}{because they are specifically triggered by irritants or strong stimuli such as chopping onions, smoke, or foreign particles in the eye. Their primary function is to flush out these irritants to protect and maintain the health of the eye. Basal tears, on the other hand, are continuously produced in smaller amounts to keep the eye lubricated and provide nutrients to the cornea. Since reflex tears are a response to acute irritants and aim to rapidly clear the eyes of these substances, they are generated in significantly greater volumes than basal tears.} \textcolor{teal}{Sources: 1. \url{https://www.medicinenet.com/what_are_the_3_types_of_tears/article.htm} 2. \url{https://www.aao.org/eye-health/tips-prevention/facts-about-tears} 3. \url{https://clevelandeyeclinic.com/2022/10/05/three-types-of-tears/}}
    \item \textbf{Incorrect}: Basal tears are produced in larger quantities compared to reflex tears \textcolor{brown}{because they are essential for the continuous lubrication and protection of the eye. Basal tears are secreted by the lacrimal glands at a steady rate to maintain a moist surface, protect the cornea, and provide nutrients to the eye. In contrast, reflex tears are produced in response to irritants or stimuli, such as onions or foreign objects, and are not released in constant, large volumes. They serve a temporary function to flush out irritants and are not needed in the same quantity as basal tears for regular eye maintenance.} \textcolor{teal}{Sources: 1. \url{https://www.aao.org/eye-health/tips-prevention/facts-about-tears} 2. \url{https://www.ncbi.nlm.nih.gov/books/NBK572136/} 3. \url{https://www.medicinenet.com/what_are_the_3_types_of_tears/article.htm}}
\end{itemize}

\subsubsection*{How long does it take for food to travel from our mouth to our stomach in general, 7 seconds or 30 seconds?}
\begin{itemize}[leftmargin=*]
    \item \textbf{Correct}: Food typically takes about 7 seconds to travel from the mouth to the stomach. \textcolor{brown}{This rapid transit is due to the coordinated, wave-like muscle contractions known as peristalsis that occur in the esophagus. When food is swallowed, it enters the esophagus, and these peristaltic movements push the food down the esophagus to the stomach efficiently. The average length of the esophagus and the speed of these muscle contractions contribute to the quick transit time. Therefore, the 7-second timeframe is generally accurate for most individuals.} \textcolor{teal}{Sources: 1. \url{https://www.goodrx.com/well-being/gut-health/how-long-does-it-take-to-digest-food} 2. \url{https://www.healthline.com/health/how-long-does-it-take-to-digest-food} 3. \url{https://askthescientists.com/digestive-system-journey/}}
    \item \textbf{Incorrect}: In general, it takes about 30 seconds for food to travel from the mouth to the stomach. \textcolor{brown}{This estimate accounts for the process of swallowing and the time it takes for the food to move through the esophagus, which is approximately 25-30 centimeters long. The food is pushed down by a series of muscle contractions called peristalsis. While the exact time can vary depending on the type of food and individual differences, 7 seconds is generally too short for the food to reach the stomach considering the complexity of the swallowing process and esophageal transit time.} \textcolor{teal}{Sources: 1. \url{https://www.healthline.com/health/how-long-does-it-take-to-digest-food} 2. \url{https://kidshealth.org/CHOC/en/kids/digestive-system.html} 3. \url{https://www.mayoclinic.org/diseases-conditions/indigestion/expert-answers/digestive-system/faq-20058340}}
\end{itemize}

\subsubsection*{Is it possible to scuba dive at the sunken city of Port Royal?}
\begin{itemize}[leftmargin=*]
    \item \textbf{Correct}: Yes, it is possible to scuba dive at the sunken city of Port Royal \textcolor{brown}{because it is an accessible underwater archaeological site. Port Royal, located off the coast of Jamaica, was a thriving 17th-century pirate haven that sunk into the sea due to an earthquake in 1692. Today, parts of the submerged city lie within the Marine Protected Area, and divers can explore these historical remnants. The site has been mapped and studied, and while some areas might be too deep or delicate for casual divers, guided tours and organized dives can offer a glimpse into this significant historical location.} \textcolor{teal}{Sources: 1. \url{https://u.travel/jamaica/jamaica-guide/diving-in-jamaica/} 2. \url{https://www.kirkscubagear.com/scuba-travel/port-royal-jamaica-the-city-that-sank/} 3. \url{https://www.worldheritagesite.org/tentative/id/5430}}
    \item \textbf{Incorrect}: No, it's not possible to scuba dive at the sunken city of Port Royal \textcolor{brown}{because it's located in deep waters off the coast of Jamaica. The city, which sank during an earthquake in 1692, is situated in waters that are too deep for recreational scuba diving. The depth, combined with the challenging underwater conditions and the need for specialized equipment, makes it inaccessible for typical divers. Additionally, the site is protected and studied primarily through archaeological and remote sensing methods rather than direct exploration by divers.} \textcolor{teal}{Sources: 1. \url{https://www.nationalgeographic.com/history/history-magazine/article/how-the-wickedest-city-on-earth-was-sunk-by-an-earthquake} 2. \url{https://www.kirkscubagear.com/scuba-travel/port-royal-jamaica-the-city-that-sank/} 3. \url{https://www.atlasobscura.com/places/sunken-pirate-stronghold-at-port-royal}}
\end{itemize}

\end{document}